\documentclass[final,5p,times,twocolumn]{elsarticle}

\usepackage{lineno}

\usepackage{color}
\usepackage{flushend}

\usepackage{graphics,picins}

\usepackage{algorithmic}
\usepackage{algorithm}

\usepackage{caption}
\usepackage{rotating}
\usepackage{amssymb,amsmath}
\usepackage{multirow}
\usepackage{verbatim}
\usepackage{lettrine}
\usepackage{flushend}
\usepackage{array,booktabs,arydshln,xcolor}
\newcommand\VRule[1][\arrayrulewidth]{\vrule width #1}

\def\draftmode{}
\ifx\draftmode\undefined
\newcommand{\mcomment}[1]{}
\else
\newcommand{\mcomment}[1]{$\Longleftarrow${\bf $<$#1$>$} }
\fi

\journal{Computer Networks}
\begin{document}


\begin{frontmatter}

\title{Bandwidth Aggregation Techniques in Heterogeneous Multi-homed Devices: A Survey}
 \author[habak]{Karim Habak}
 \ead{karim.habak@cc.gatech.edu}

 \author[kharras]{Khaled A. Harras}
 \ead{kharras@cs.cmu.edu}

 \author[youssef]{Moustafa Youssef}
 \ead{moustafa.youssef@ejust.edu.eg}

 \address[habak]{School of Computer Science, College of Computing, Georgia Institute of Technology, Atlanta, Georgia.}
 \address[kharras]{Computer Science Department, School of Computer Science, Carnegie Mellon University Qatar, Doha, Qatar}
 \address[youssef]{Wireless Research Center, Egypt-Japan University of Science and Technology (E-JUST) and Alexandria University, Alexandria, Egypt.}



\begin{abstract}
The widespread deployment of various networking technologies, coupled with the exponential increase in end-user data demand, have led to the proliferation of multi-homed, or multi-interface enabled, devices. These trends drove researchers to investigate a wide spectrum of solutions, at different layers of the protocol stack, that utilize available interfaces in such devices by aggregating their bandwidth. In this survey paper, we provide an overview and examine the evolution of bandwidth aggregation solutions over time. We begin by describing the bandwidth aggregation problem. We then investigate the common features of proposed bandwidth aggregation systems and break them down into two major categories: layer-dependent and layer-independent features. Afterwards, we discuss the evolution trends in the literature and discuss some open challenges requiring further research. We end the survey with a brief presentation of related work in tangential research areas.
\end{abstract}

\begin{keyword}
Bandwidth aggregation \sep multiple network interfaces \sep throughput \sep optimization \sep multihoming
\end{keyword}
\end{frontmatter}

\section{Introduction}
\lettrine{W}{ith} the continuous advancement of wireless technologies,
decreasing cost of electronics, and the heterogeneity of network access
technologies, it is becoming the norm nowadays to have multi-interface enabled
devices, also known as multi-homed devices. Having such devices provides an
opportunity for leveraging their interfaces to meet the increasing user demand
for bandwidth and handle the increasing Internet traffic sizes
\cite{gebert2012internet,labovitz2011internet,attReport}. Unfortunately, the
state-of-art operating systems fail to utilize the true potentials of these
interfaces. For instance, the vast majority of these operating systems, such as
Windows, Linux, and Mac OS, typically assign all the applications' traffic to
one of the available interfaces, {\it even if more than one is connected to the
Internet.} On the other hand, the recent deployment of MPTCP on a subset of
mobile devices enables them to utilize the available interfaces {\it while
using a limited set of applications and communicating a limited number of
servers}. Overall, the failure of state-of-art operating systems to leverage
the available interfaces in multi-homed mobile devices leads to
under-utilization of available bandwidths, waste of potential connectivity
resources, and dissatisfaction of users.

The fundamental approach for leveraging multiple network interfaces on
multi-homed mobile devices is aggregating the bandwidth available on each of
these interfaces. We define bandwidth aggregation as the ability to leverage
the available network interfaces to increase the bandwidth for users and
applications. Over the past decade, a large body of research has emerged to
enable bandwidth aggregation on multi-homed devices. To achieve their main goal
of enhancing the user experience, they provide solutions for two major of
challenges: (1) core aggregation challenges and (2) Internet integration
challenges. First, core aggregation challenges refer to the set of challenges
that come with bandwidth aggregation even when designed as a clean slate.
Examples of these challenges include estimating network interface
characteristics, and scheduling data across different interfaces.  Second,
Internet integration challenges refer to the set of challenges introduced by
the current design of the Internet and its protocol layers. For instance, the
vast majority of current Internet applications use TCP over IP as a result of
historical decisions tightly linking both protocols. These decisions led to the
development of systems and applications that ultimately expect to run on a
single network interface, thus identifying communication end points by a single
IP address at each end. As a result, many solutions have been developed and
implemented at different layers of the TCP/IP protocol stack to solve the two
challenges above and work around the current Internet design and
characteristics.

In this paper, we survey the current state-of-the-art solutions for the
bandwidth aggregation problem. We categorize, study, and share the various
solutions implemented at different layers of the protocol stack. Solutions
implemented in the same layer usually share common goals, challenges, and
possess what we denote as \textit{layer-dependent features}. On the other hand,
there are other common features shared between all the solutions regardless of
the layer they are implemented in. These \textit{layer-independent features}
include estimating the interface and applications characteristics, scheduling,
and network support and communication model. In addition, we analyze the
evolution of the solution space and discuss the open challenges and new trends.
Although a previous survey for bandwidth aggregation has been conducted
\cite{OldSurvey}, we are taking a fundamentally different approach while
surveying this area which enables us to (1) reveal new relations between the
existing solutions, (2) build a framework for developing and deploying a
bandwidth aggregation system, and (3) discover new challenges for researchers
to address. In addition, the existing survey lacks many key papers added in
this survey. It also only focuses on scheduling and packet reordering
challenges which is not sufficient for truly characterizing the efficiency of
the bandwidth aggregation solutions. It also neglects deployment challenges,
identifying different layer-dependent and independent features, shedding light
on the chronological evolution of related literature over the past decade, and
does not suggest enough open research challenges.

In addition to the core area surveyed in this paper, we also identify three
tangential research areas that share some characteristics with bandwidth
aggregation. First, multi-path routing addresses problems resulting from having
multiple paths to a given destination \cite{survey}. Second, resource
aggregation in computer sub-systems, investigates obtaining higher performance
by aggregating other computer resources such as hard disks \cite{LS2}. Third,
multiple network interfaces can be utilized for minimizing energy consumption
\cite{Wiff20}, handling mobility \cite{Intentional2}, controlling/redirecting
traffic, using multiple channels \cite{Intentional10}, or avoiding primary
users in cognitive radio networks \cite{LAUNCH}.

The rest of the paper is organized as follows. Section \ref{LDF} discusses the
layer-dependent features. In Section \ref{LIF}, the layer-independent features
are presented. Section \ref{Evol} analyzes the area evolution over time as well
as the open research challenges. In Section \ref{RW}, we discuss other
tangential areas that have similar challenges and solutions. Finally, the paper
concludes in Section \ref{con}.

\section{Layer-dependent Features}
\label{LDF}
\begin{figure}[!t]
  \centerline{\includegraphics[width = 1.05\columnwidth]{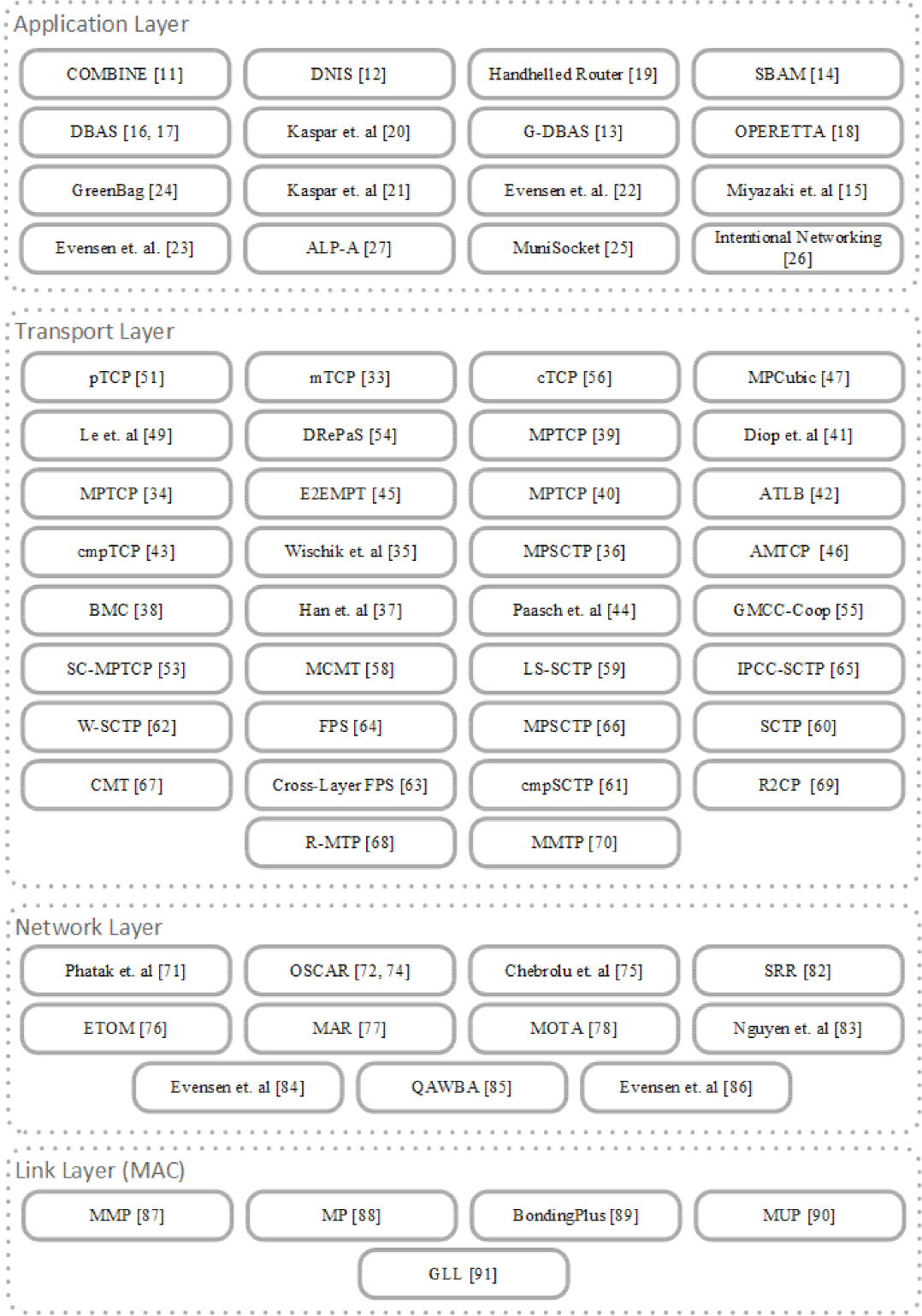}}
  \caption{Bandwidth aggregation solutions and their corresponding location in the protocol stack}
  \label{LDFFigure}
\end{figure}

A number of solutions have been proposed for the bandwidth aggregation problem
on multi-homed devices. These solutions, shown in Figure \ref{LDFFigure}, are
implemented at different layers of the protocol stack. Typically solutions
implemented in a specific layer of the protocol stack share address similar
challenges and face some limitations depending on their selected implementation
layer. In this section, we discuss the layer-dependent features of the
different proposed techniques. We end this section with providing a brief
summary in Table~\ref{Summary-Table}. We defer the discussion of the
layer-independent features to the next section.

\subsection{Application Layer Solutions}
\label{sec:LDFAPP}

Apart from the multi-homed aware applications, which utilize available
interfaces for their own benefit while lacking global efficiency and interface
load awareness \cite{combine}, application layer solutions are mainly
implemented as a middleware between the application and transport layers. This
middleware takes the responsibility of handling multiple interfaces and
distributing the different application data across them. We break down these
solutions into two main categories: (1) transparent middleware and (2)
non-transparent middleware.

\subsubsection{Transparent Middleware} 
Transparent middlewares are designed to work seamlessly with current
applications and enable them to make use of multiple available interfaces. In
such cases, the middleware implements the same interface that the transport
layer provides to the application. This middleware also guarantees the same
semantics provided by the traditional transport layer. Therefore, transparent
middlewares have the advantage of ease of deployment and backward compatibility
as they do not require changes to current applications. However, they are
usually more complex to implement compared to the non-transparent middlewares.

Solutions that implement a transparent middleware on top of a reliable
transport layer protocol, such as TCP, must guarantee reliability.  However,
working on multiple interfaces may produce out-of-order delivery. Therefore,
such solutions need to provide mechanisms for correct in-order packet delivery.
To handle this, some solutions provide connection-level granularity scheduling
where all the packets belonging to the same TCP connection will traverse the
same interface while different connections can use multiple interfaces
concurrently \cite{DNIS,G-DBAS}. This approach enables these solutions to
operate with the conventional Internet architecture and servers. Other
solutions, however, implement a packet-reordering technique at both endpoints
\cite{SBAM,miyazaki2012evaluation}. Although these solutions show a great
potential to increase system performance, upgrading the servers becomes a
formidable barrier preventing their large-scale deployment.

To minimize the widespread deployment cost of the transparent middlewares,
researchers avoid relying on upgrading the servers to enable bandwidth
aggregation. Habak et al. use connection-level granularity scheduling as the
default mode of operation while leveraging optional server modifications to
further enhance performance \cite{DBAS,DBAS-ComNets,OPERETTA}. Moreover, Sharma
et al. rely on a proxy server to hide the effects of using multiple interfaces
at the client side form legacy servers  \cite{TMCMCC}. In this case,
aggregating bandwidth occurs between clients and proxy servers. 

Exploiting application-layer protocol functionalities to aggregate the
bandwidth of different interfaces without network or server support has also
gained researchers attention. For instance, Some researchers exploit the
availability of HTTP range queries for bandwidth aggregation
\cite{UsingHTTP,EnhanceVideo}. In particular, they break an HTTP stream into
several pieces and open a new TCP connection to get each piece separately,
using an HTTP range query requesting each piece. To aggregate the interfaces
bandwidth, they distribute these TCP connections across available network
interfaces. Therefore, they are able to exploit these interfaces while using
HTTP protocol with the support of range queries at the server side. On the
other hand, motivated by the advent of dynamic adaptive streaming over HTTP
(DASH) and its adoption by Youtube, Hulu, and Netflix, Evensen et al. build on
the idea of using application-layer protocol functionalities for bandwidth
aggregation by utilizing the DASH request-response communication model
\cite{QualityAdaptive,QualityAdaptive2}. In this case, instead of issuing
different video-segment requests using the same interfaces, they distribute
these requests across the different interfaces to enhance system performance
and enable users to stream higher quality videos. Although, GreenBag
\cite{GreenBag} adopts the idea of distributing video requests across
interfaces, they focus on minimizing the time of video playback along with
energy consumption, while overlooking the ability to stream higher quality
videos.

\subsubsection{Non-transparent Middleware}
Non-transparent middlewares are designed to provide applications with simple
APIs for using multiple network interfaces. Therefore they require
modifications to the applications in order to make use of available interfaces.
In such cases, the middleware is able to modify some terms in the agreement
between the application and the transport layer. These modifications aim to
enhance the overall system performance as well as provide the application with
what it needs. For example, MuniSocket changes the agreement between the
application and transport layer such that it supports reliability upon request
by the applications \cite{MuniSocket}. Basically, it uses UDP to transmit
packets, and engages its implemented reliability mechanism if requested by the
application. Intentional networking, however, changes the agreement to
guarantee the ordering of data only upon a request from the application
\cite{Intentional}. The application defines its own ordering constraints and
intentional networking guarantees satisfying these constraints. This guarantee
is achieved by defining IROBs (Isolated Reliable Ordered Bytestreams) as the
unit of data transmission defined by the application. Intentional networking
gives each IROB a unique identifier. While creating an IROB, an application can
specify a list of IROBs that have lower identifiers that must be received prior
to receiving this IROB. Although ALP-A \cite{ALP-A} exploit the HTTP protocol's
features to enable bandwidth aggregation, they rely on applications to specify
their required level of quality of experience (QoE) by defining a deadline for
each HTTP request.

\subsection{Transport Layer Solutions}
\label{sec:Transport}
As shown in Figure \ref{LDFFigure}, many bandwidth aggregation solutions
naturally lie in the transport layer. We classify these solutions into three
categories: (1) extending widely deployed protocols (e.g. TCP), (2) utilizing
multi-homing support in existing protocols (e.g. SCTP) and (3) designing new
transport protocols.

\subsubsection{Extending Widely Deployed Protocols}
Since TCP has been the dominating protocol for Internet traffic for the past
decades, a lot of work in the literature focuses on extending it to support
transmission over multiple network interfaces. As a result, there exists many
TCP extensions that leverage multiple interfaces.  These protocols usually
address issues that hinder TCP from utilizing the available interfaces in
parallel such as (1) reordering packets belonging to different sub-flows, (2)
scheduling packets across multiple interfaces, (3) node identification, and (4)
congestion control mechanisms.

Packet recording is usually performed using a global reordering buffer at the
receiver. We defer the discussion about packet scheduling to the
layer-independent features in Section \ref{Scheduling}. On the other hand,
since a TCP connection is defined by the source and destination ports and IP
addresses, all TCP extensions enable the source and/or the destination to have
multiple IPs. These IPs are exchanged between the two endpoints in the
beginning and during the lifetime of a connection. The rest of this section
focuses on the TCP congestion control mechanisms because they are the core
functionality of TCP and unique to the transport layer.

\label{sec:CongCon}
\textit{\textbf{- Congestion Control:}} Congestion control is the key component
of the TCP protocol. This component introduces most of the TCP's important
features such as fairness and ability to utilize the available bandwidth in the
underlying network without overwhelming it. Therefore, several researchers
develop congestion-control mechanisms to enhance TCP performance in different
situations such as using wireless links \cite{WTCP}, utilizing satellite links
\cite{STP}, and operating on top of high-bandwidth networks
\cite{compoundTCP,cubicTCP}. Unfortunately, these mechanisms are designed to
utilize only one underlying path, and thus, they assume that packets are
supposed to arrive in-order. Therefore the arrival of unexpected packet
(out-of-order packet arrival) indicates packet loss happened due to
network-congestion, which requires decreasing the transmission rate.
Furthermore, these congestion control mechanisms, also, use only one timeout
variables to detect packet loss due to severe network congestion. Due to these
assumptions, using any of these mechanisms in the TCP protocol makes it
incapable of working on multiple network interfaces.

Running the TCP protocol on multiple heterogeneous interfaces is not only
inefficient, but also it may cause performance degradation. While using
heterogeneous interfaces, out-of-order packet delivery becomes the norm because
of differences in bandwidths and latencies between the interfaces. This
out-of-order packet delivery causes unnecessary shrinking of the congestion
window causing drop in transmission data rates. Furthermore, interface
heterogeneity may lead to unnecessary timeouts, which lead to severe dropping
of the congestion window size. In addition, even if congestion was detected
that affects an interface, the excessive decrease of the congestion window
leads to decreasing the transmission rate on other interfaces as well. In many
cases, running TCP on top of multiple interfaces achieves lower performance
compared to running it on top of only one of these interfaces
\cite{hsieh2002ptcp}. Therefore, solutions that extend TCP to support
multi-interface communication modify its congestion control mechanism.  

\begin{figure}[!t]
  \centerline{\includegraphics[width=\columnwidth]{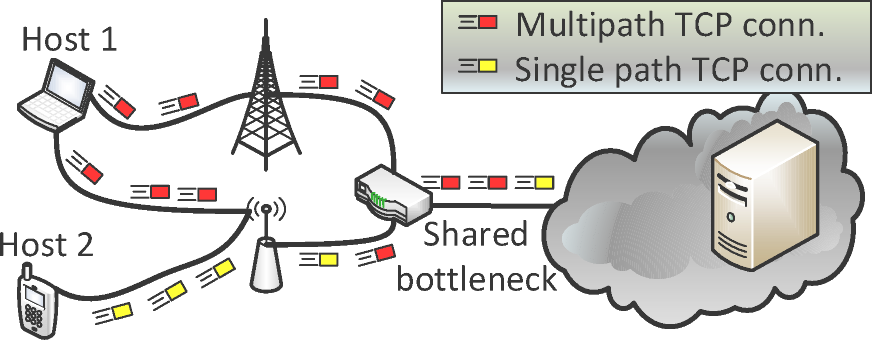}}
  \caption{Unfair bandwidth distribution between multipath TCP and regular TCP. Host~1 achieves double the bandwidth achieved by Host~2 because It uses multi-path variant of TCP, in which sub-flows control the congestion independent of each others. Many solutions  were developed to address this problem \cite{mTCP,shamszaman2013feasibility,wischik2011design,raiciu2012hard,han2004overlay,honda2009multipath,peng2013multipath}.}
  \label{fairness}
\end{figure}

To efficiently utilize the available interfaces, TCP extensions modify the
congestion control mechanism in different ways. Many extensions propose
applying a congestion control mechanism to each sub-flow independent of the
other sub-flows, where a sub-flow is the portion of the connection that is sent
over the same path. This technique implicitly decouples the congestion control
problem from the packet reordering one. However when it comes to building their
protocols, these extensions adopt different congestion control mechanisms based
on the designed goals and the target usage scenario.

For the sake of simplicity, many protocols use the same congestion control
mechanism while deploying it on each sub-flow, independently. For example, many
protocols deploy the standard congestion control mechanism of TCP on each
sub-flow
\cite{MPTCP,QoSTCP,hasegawa2007,cmpTCP,paasch2012exploring,E2EMPT,AMTCP}. In
contrast, to achieve best utilization of high bandwidth links, MPCubic
\cite{multipathCubic} uses a cubic congestion control mechanism \cite{cubic}
and Le et al. \cite{BinomialCC} use a binomial congestion control mechanism
\cite{bansal2001binomial} for each of their sub-flows.  Meanwhile, to
efficiently operate on top of heterogeneous interfaces, pTCP \cite{pTCP} allows
the use of a different TCP variant, e.g. \cite{STP,TCP-Westwood,WTCP}, for each
interface depending on its characteristics. Each of them runs its own
congestion control mechanism independently and handles its interface
characteristics accordingly. For example, using pTCP on a host that is equipped
with a wireless WAN (WWAN) network interface and satellite interface will make
it use WTCP \cite{WTCP} for WWAN interface and STP \cite{STP} for the satellite
interface.  Although these approaches enable TCP to efficiently utilize
multiple interfaces in many scenarios, they lose its fairness property in case
of having shared bottlenecks. Figure \ref{fairness} shows a scenario, in which
running a congestion control mechanism on each sub-flow independently leads to
unfair distribution of the available bandwidth over running connections. In
particular, this figure shows that multipath TCP \cite{MPTCP} obtains double
the bandwidth obtained by TCP.

To achieve fair bandwidth distribution in case of having shared shared
bottlenecks while achieving the maximum utilization of the available
interfaces, many researchers propose tweaks to the congestion control schemes.
mTCP uses the standard congestion control mechanism of TCP on each sub-flow
independent of the other sub-flows \cite{mTCP}. However, when mTCP detects that
two or more sub-flows share their bottleneck, It merges them together as one
sub-flow and uses one congestion window for them. To detect that two interfaces
has a shared bottleneck, they use the correlation between fast retransmission
timestamps on both interfaces. Unfortunately, this approach takes seconds
(maximum 15 seconds) to detect the existing shared bottlenecks. Therefore, mTCP
achieves the fairness goals for only long connections. To overcome this
problem, many protocols couple the congestion control mechanisms, which
controls the rates on the running sub-flows
\cite{shamszaman2013feasibility,wischik2011design,raiciu2012hard,han2004overlay,honda2009multipath,peng2013multipath,SC-MPTCP,DRePaS}.
Generally, these approaches limit the growth in the congestion window of each
sub-flow based on various parameters such as the sum of the congestion windows
of all sub-flows, the delay and congestion correlation with other sub-flows,
and estimates of a competing-TCP throughput.

GMCC-Coop \cite{GMCC-Coop} extends the fairness definition to be suitable for
bandwidth sharing scenarios. Its fairness definition (1) enables TCP and MPTCP
connections originated in a sharing node achieve more bandwidth compared to the
MPTCP sub-flows it relays, and (2) forces MPTCP to achieve the same bandwidth
of TCP under similar loss conditions. GMCC-Coop achieves this by limiting the
rate of increase of the congestion window size specially for the sub-flows
relayed through a bandwidth sharing neighbor.

Instead of running different congestion control mechanism at each sub-flow,
cTCP modifies TCP's congestion control mechanism to deal with multiple
interfaces\cite{cTCP}. The protocol uses a database at the sender to keep track
of all the packets that have been sent but not acknowledged and the interface
used for sending each of these packet. In this protocol, the receiver sends an
acknowledgement on the path used by the received packet, which triggered this
acknowledgement to be sent.  When an in-sequence packet acknowledgement is
received, the sender deletes all the packets up to the one being acknowledged
from the database. However, when a duplicate acknowledgment (ACK) is received
with a packet sequence number equal to $n$, this highlights that the receiver
is still waiting for the $(n+1)^{\textrm{th}}$ packet, while receiving
subsequent packets to that. If the duplicate ACK is received over the same path
used for sending the $(n+1)^{\textrm{th}}$ packet, this is considered a real
duplicate acknowledgement due to packet loss. Otherwise, a race condition has
occurred resulting in duplicate ACKs created due to differences in path
characteristics. This intelligent handling of acknowledgments reduces the
packet reordering problem.

\subsubsection{Utilizing Multi-homing Support in Existing Protocols}
The stream control transmission protocol (SCTP) is one of the protocols that
researchers heavily investigated while proposing solutions for the bandwidth
aggregation problem due to its inherent design that supports multi-streaming
and multi-homing \cite{SCTPRFC}. Fortunately, SCTP allows data to be
partitioned into multiple streams. Each of these streams independently delivers
its portion of the data to the application running at the receiver. This means
that the loss of a data chunk belonging to a certain stream only affects the
delivery within that stream. This feature prevents the head-of-line blocking
problem that can occur in TCP, since TCP only supports single streams. In
addition, multi-homing also allows a single SCTP endpoint to support multiple
IP addresses. SCTP multi-homing support, however, is only for redundancy. A
single address is chosen as the primary address, which is the destination
address for all data chunks during normal transmission. These characteristics
of the SCTP protocol encourages bandwidth aggregation researchers to work on
extending it in order to exploit the available interfaces in parallel.

Similar to extending TCP, the work extending SCTP focuses on reordering packets
belonging to different sub-flows or streams, scheduling packets across the
different interfaces, and developing appropriate congestion control mechanisms.
To discuss the detailed characteristics of these extensions, we categorize them
into two main categories: (1) application-assisted aggregation, where
applications provide some assistance to the SCTP extension such as defining
their different streams or setting relations between their data unites, and (2)
application-oblivious aggregation, where the existence of multiple interfaces
is hidden from the applications.

\textit{\textbf{- Application-Assisted Aggregation:}} MCMT  is an extension to
the SCTP protocol that uses an application-assisted aggregation approach
\cite{huang2010design}. This extension utilizes the multi-streaming feature in
the SCTP protocol to solve the previously mentioned challenges. It gives the
applications the responsibility to define their streams. Then, it adopts a
path-oriented multi-streaming scheduling technique in which the packet that
belongs to the same stream utilizes the same path. For example, an application
streaming a video from an MCMT-enabled server is responsible for dividing its
data into two streams. The first stream is used to transmit the video while the
second is used for transmitting the related audio. In this case MCMT will
transmit all video packets using the same path which may be different from the
path used for transmitting the audio data. Assigning streams to paths is the
task of the scheduler which we present in Section \ref{Scheduling}. In
addition, the application may further divide the video or audio into multiple
streams to enhance performance. However, it will carry the overhead of
reordering packets and applying its own reliability requirements on each
stream.

\textit{\textbf{- Application-Oblivious Aggregation:}} Many extensions to SCTP
focus on developing an application-oblivious protocol that seamlessly aggregate
the available bandwidth of the network interfaces. Therefore, they have to
maintain the same contracts between applications and the SCTP protocol. Hence,
they have to provide solutions to the previously mentioned problems. In
particular, packet reordering is implemented by having a global reordering
buffer for each stream. This buffer is used in case there are no requests for
out-of-order packet delivery from the applications. In Section
\ref{Scheduling}, we address the scheduling mechanisms used in detail.

Because of the great similarities in adopted congestion control protocols
between an SCTP stream and a TCP connection, SCTP extensions attempt to address
the same challenges discussed in Section~\ref{sec:CongCon} while extending
their congestion control mechanism. Similar to extending the congestion control
of TCP, many solutions implement the standard SCTP stream congestion control
mechanism at each sub-stream independent of the other sub-streams
\cite{LS-SCTP,SCTP,cmpSCTP,W-SCTP,CFPS,FPS,IPCC-SCTP}. On the other hand, other
SCTP extensions deploy a congestion control mechanism on the whole stream
\cite{MPSCTP,CMTSCTP}. In this case, they change the techniques of detecting
congestion and update the congestion window to be suitable for running over
more than one interface. They change the fast retransmission technique such
that a retransmission and a congestion windows decrease are triggered by
out-of-order packet delivery for the packets that utilize the same path. Hence,
they store the path used to send each packet, thus, when selective
acknowledgement (SACK) identifies a gap in certain path, it triggers fast
retransmission and congestion window update.

\subsubsection{Designing New Protocols}
While protocols presented in this section are newly designed, some of these
protocols maintain the same application-transport contract of TCP due to its
wide spread use and deployment.

Magalhaes et al. propose the R-MTP transport layer protocol which uses
retransmission-based reliability and gap-detection for identifying losses
\cite{R-MTP}. The sender is notified that frames have arrived at the receiver
by acknowledgements. R-MTP's gap-detection relies on selective acknowledgment.
In order to control the network congestion, R-MTP introduces a new congestion
control mechanism: The receiver is the entity responsible for detecting
congestion by monitoring the delay jitter calculated from the difference in
time between every two consecutive packets and its mean value. This mean value
is calculated from the rate which the sender and receiver agreed on. The idea
is that, in the case of no congestion, the long term jitter should be close to
zero. Hence, the increase in the delay jitter indicates network congestion.
This congestion control technique is applied to each path independently.

RCP is another example of newly designed protocols maintaining TCP's contract
between the application and the transport \cite{RCP}. RCP is a receiver-centric
transport protocol designed to avoid the TCP limitations. It is implemented to
deal with one network interface while keeping in mind the ease of extension to
support multiple network interfaces. The authors extend this protocol to
support communication through multiple network interfaces by proposing R$^2$CP.
In RCP, reliability is implemented by making the receiver request data from the
sender instead of acknowledging the data. They define two types of requests: 1)
cumulative request, which is used in order to request new data, and 2) pull
request, which is used to request packet retransmissions. Flow control in RCP
is much easier than that of TCP where a receiver only sends requests when it
has free space in its buffer. Congestion control is similar to TCP's except for
it being receiver-centric. R$^2$CP applies this congestion control in each
sub-flow to support additional interfaces.


MMTP \cite{MMTP} is designed to utilize the available interfaces to achieve the
demanding multi-media's bandwidth requirements. This protocol is designed to
have the frame received before its deadline and avoid wasting network resources
in sending frames that are going to be useless due to late arrival.

\subsection{Network Layer Solutions}
Network layer solutions target maintaining adopted and deployed transport layer
protocols and allowing them to work efficiently on different network interfaces
by making modifications to the network layer. Due to the TCP's popularity, most
of network layer solutions usually use it as the target transport layer
protocol. These solutions  consequently address three main issues that prevent
TCP from achieving high performance while running on multiple interfaces: (1)
breaking TCP's connection semantics, (2) congestion misprediction, and (3)
round trip time (RTT) estimation.

\subsubsection{Breaking TCP Connection Semantics} Since each interface has its
own IP address, distributing packets that belong to the same connection over
multiple interfaces breaks the TCP connection semantics that identifies a
connection by the source and destination IPs and port numbers. To address this
problem, network layer solutions hide the usage of multiple IPs from the
running TCP protocol. For instance, Phatak et al. use IP-in-IP encapsulation to
hide the usage of multiple IPs from TCP \cite{Novel}.  In this case, the source
and destination open a TCP connection with one IP for each of them. These IPs
are used for all packets to/from the transport layer. When a packet is sent
using another interface, or sent to an interface other than the one agreed on
during connection establishment, the packet with the agreed upon IP from the
transport layer is encapsulated in another packet whose header contains the
actual interface IP. The network layer at the destination extracts this packet
and forwards it to the destination transport layer. Fortunately, performance
evaluation showed that the encapsulation overhead is negligible \cite{Novel}.
To achieve the same goal, OSCAR uses network address translation (NAT) instead
of IP-in-IP encapsulation \cite{OSCAR,OSCAR-MOBI,OSCAR-MASS}. In this case,
OSCAR replaces the source and destination IPs at the sender with the used IPs
for transmission. Upon receiving a packet, the receiver reverses the source and
destination IPs by replacing them with the negotiated ones before giving the
packet to TCP.  Although the encapsulation and the NATing techniques show
efficiency, implementing them requires updating the network layer at the
endpoints.

To ease deployment, many solutions attempt to avoid upgrading servers while
proposing solutions that hide using multiple IPs from TCP. For example,
Chebrolu et al. rely on a proxy to hide client multiple IPs from the server
\cite{networkTCP}. This proxy interacts with servers using a single IP address
and is aware of the client's multiple IPs while communicating with it. The
solution adopts IP-in-IP encapsulation between the proxy and the client to hide
the client IPs from the running TCP connection at the client side. ETOM,
however, adopts a different architecture which consists of a client, a server,
a proxy server and a router equipped with multiple interfaces \cite{ETOM}. It
splits a connection between the client and the server in three parts: (1) A
normal TCP connection between the client and the router, (2) A normal TCP
connection between the proxy server and the server, (3) multiple TCP
connections between the router and the proxy server such that each of these
connections utilizes only one path. In this case, they do not need to hide the
used IPs from the running TCP connections. Furthermore, MAR \cite{MAR} uses a
similar architecture with the following differences: (1) communication between
the router and the proxy is not limited to using multiple TCP connections and
(2) the proxy is optional to minimize the deployment cost.  In the absence of a
proxy, MAR provides a per-TCP connection mode of operation, in which each
connection is assigned only to one interface but different connections can be
assigned to different interfaces. MOTA, on the other hand, adopts a special
case of connection-oriented scheduling in which all the application load is
assigned to only one network, while different applications can be assigned to
different networks \cite{mota}.

We highlight that the problem of breaking TCP connections due to using multiple
IP addresses appears in multiple contexts other than multi-interface bandwidth
aggregation. For example, mobile devices change their IP addresses while
moving, thus, handling user mobility and maintaining active connections also
deal with this problem \cite{MIP}. Therefore current state-of-the-art in these
research areas can provide bandwidth aggregation researchers with mechanisms to
solve this problem such as host identification protocol (HIP) \cite{HIP}. In
addition, although some clean slate Internet architectures provide solutions to
this problem \cite{XIA}, they still introduce new set of challenges for
bandwidth aggregation researchers.

\begin{table*}
\centering
  \begin{tabular}{ | c || c || c |}
    \hline
    \multirow{6}{*}{\parbox{1.6cm}{\centering Application Layer}} &
    In-application & COMBINE\cite{combine} \\\cline{2-3}
   
    & \multirow{3}{*}{\parbox{2.5cm}{\centering Transparent Middlewares}} &
    DNIS \cite{DNIS} \;\; Handhelled Router\cite{TMCMCC} \;\;
    SBAM\cite{SBAM} \;\;DBAS\cite{DBAS,DBAS-ComNets}\\ &  &  Kaspar et al.
    \cite{UsingHTTP} \;\; G-DBAS\cite{G-DBAS} \;\; OPERETTA\cite{OPERETTA}
    \;\; GreenBag\cite{GreenBag} \;\; Kaspar et al.\cite{EnhanceVideo} \\ &
    &  Evensen et al. \cite{QualityAdaptive} \;\; Miyazaki et
    al.\cite{miyazaki2012evaluation} \;\; Evensen et
    al.\cite{QualityAdaptive2} \;\; ALP-A\cite{ALP-A}\\\cline{2-3}
 
    & \multirow{2}{*}{\parbox{2.5cm}{\centering Non-transparent Middlewares}} &
    \multirow{2}{*}{MuniSocket\cite{MuniSocket} \;\; Intentional
    Networking\cite{Intentional}}\\ &  &  \\ \hline

    \multirow{8}{*}{\parbox{1.6cm}{\centering Transport Layer}} &
    \multirow{4}{*}{\parbox{2.5cm}{\centering Extending Widely Deployed
    Protocols}} & pTCP\cite{pTCP} \;\; mTCP\cite{mTCP} \;\; cTCP\cite{cTCP}
    \;\; MPCubic \cite{multipathCubic} \;\; Le et al. \cite{BinomialCC} \;\;
    DRePaS \cite{DRePaS} \\ & & MPTCP \cite{peng2013multipath} \;\;  Diop et
    al. \cite{QoSTCP} \;\; MPTCP \cite{shamszaman2013feasibility} \;\; E2EMPT
    \cite{E2EMPT} \;\; MPTCP \cite{MPTCP} \\ &  & ATLB \cite{hasegawa2007}
    \;\; cmpTCP \cite{cmpTCP} \;\; Wischik et al. \cite{wischik2011design}
    \;\; MPSCTP \cite{raiciu2012hard} \;\; AMTCP \cite{AMTCP}\\ &  &
    BMC\cite{honda2009multipath} \;\; Han et al. \cite{han2004overlay} \;\;
    Paasch et al. \cite{paasch2012exploring} \;\; GMCC-Coop \cite{GMCC-Coop}
    \;\; SC-MPTCP \cite{SC-MPTCP} \\\cline{2-3}
    
    & \multirow{3}{*}{\parbox{2.5cm}{\centering Utilizing Existing Support of
    Multi-homing}} & MCMT \cite{huang2010design} \;\; LS-SCTP\cite{LS-SCTP}
    \;\; IPCC-SCTP \cite{IPCC-SCTP} \\ &  & W-SCTP \cite{W-SCTP} \;\; FPS
    \cite{FPS} \;\; MPSCTP\cite{MPSCTP} \;\; SCTP \cite{SCTP} \;\; CMT
    \cite{CMTSCTP}\\ &  & Cross-Layer FPS \cite{CFPS} \;\; cmpSCTP
    \cite{cmpSCTP}  \\\cline{2-3}
    
    & \multirow{1}{*}{\parbox{2.5cm}{\centering New Protocols}} &
    R$^2$CP\cite{RCP} \;\; R-MTP\cite{R-MTP} \;\; MMTP \cite{MMTP}\\ \hline
    
    \multirow{3}{*}{\parbox{1.6cm}{\centering Network Layer}} &
    \multirow{2}{*}{\parbox{2.5cm}{\centering For TCP}} & Phatak et al.
    \cite{Novel} \;\; OSCAR \cite{OSCAR,OSCAR-MASS} \;\; Chebrolu et al.
    \cite{networkTCP} \;\; SRR\cite{SRR} \\ &  &  ETOM \cite{ETOM} \;\; MAR
    \cite{MAR} \;\; MOTA \cite{mota} \;\;  Nguyen et al.
    \cite{nguyen2013tcp}\\\cline{2-3}
    
    & For UDP/Others &  Evensen et al. \cite{MultipathUDPBandwidth} \;\;
    QAWBA \cite{QAWBA} \;\; Evensen et al. \cite{NetworkUDP}\\ \hline

    \multirow{2}{*}{\parbox{1.6cm}{\centering Link Layer (MAC)}} & Wired
    Networks&  MMP\cite{MPPP} \;\; MP\cite{PPPMul}  \;\;
    BondingPlus\cite{Bondingplus}  \\\cline{2-3}
    
    & Wireless Networks& MUP\cite{MUP} \;\; GLL\cite{GLL} \\
   	
    \hline
  \end{tabular}
  \caption{Layer based clustering of bandwidth aggregation solutions.}
\label{Summary-Table}
\end{table*}
 \subsubsection{Congestion Misprediction} When running TCP on
top of multiple network interfaces that vary in terms of delay and available
bandwidth, out-of-order packet delivery becomes the norm, leading to
unnecessary drop in the congestion window and the transmission rate of TCP
(Section~\ref{sec:CongCon}). Therefore, hiding the out-of-order packet delivery
from TCP is a critical feature of network layer bandwidth aggregation
techniques. Some solutions solve this problem by implementing packet reordering
within the network layer
\cite{ETOM,networkTCP,OSCAR,OSCAR-MASS,OSCAR-MOBI,nguyen2013tcp}. Instead of
delivering out-of-order packets to TCP, they buffer out-of-order packets in the
network layer until the preceding packets arrive. Although this approach hides
the out-of-order packet delivery from TCP and, thus, avoids the unnecessary
shrinking of the congestion window, it should be carefully implemented since it
may result in detecting packet loss only via timeout at the sender resulting in
severe congestion window drops. Therefore, these solutions implement packet
loss detection techniques at the network layer to avoid timing out on the lost
packets. When a loss is detected at the network layer (e.g. by setting a
threshold on the packet's waiting time in the reordering queue), the network
layer forwards the received out-of-order packets to TCP to trigger duplicate
ACKs or selective ACKs (SACKs) which is used to detect the loss quickly and
more importantly to avoid a timeout event, which is more costly than a
duplicate ACKs or SACKs.

We note that MAR offers the architecture in which a protocol is used between a
multi-homed router and the optional proxy\cite{MAR}. In order to achieve high
performance, this protocol must be carefully designed such that it handles
reordering and hides it from the end points. In addition, another approach to
avoid this issue is using connection oriented scheduling which is adopted by
MAR in case of no proxy, and OSCAR while communicating with legacy servers.

\subsubsection{Round trip time estimation technique} As a result of using
multiple interfaces, each connection can go through multiple paths that vary in
their behavior, including the round trip time. In addition, reordering affects
the calculation of round-trip time (RTT) estimation and hence determining the
right value for the retransmission timeout timer (RTO). Hence, Phatak et al.
\cite{Novel} study the effect of distributing the data across the different
network interfaces on the RTT and RTO estimation. They address this problem by
building a mathematical model to avoid the negative effects of errors in
estimating the RTT and determining the RTO and take this into account in their
scheduling decision as we discuss in Section \ref{Scheduling}. Others handle
this problem by implementing reordering at the network layer
\cite{ETOM,networkTCP,OSCAR,OSCAR-MASS,nguyen2013tcp}. This reordering delays
the packets from the fast paths waiting for previous packets to arrive which
were sent on the slow paths. This makes RTO and RTT estimations bound by the
slowest path.  On the other hand, connection-oriented scheduling provides
another solution for this issue \cite{MAR,OSCAR,OSCAR-MASS}.

\subsection{MAC Layer Solutions} MAC layer solutions are the first bandwidth
aggregation solutions to emerge to address problems such as providing enough
communication bandwidth between database servers. These solution are limited to
work in scenarios where devices are directly connected through multiple links.

In wired networks, there are many MAC layer protocols designed to aggregate the
bandwidth of multiple links connecting two devices. Some of these protocols
utilize identical interfaces \cite{PPPMul,MPPP}. On the other hand, BondingPlus
aggregates multiple Ethernet links by introducing a bonding layer, below the
network layer, responsible for distributing packets across these Ethernet links
\cite{Bondingplus}. It also extends the ARP protocol to implement ARP+ which
enables maintaining multiple MAC addresses for the same IP address.

In wireless networks, protocols are designed to aggregate the bandwidth of two
or more radio interfaces tuned to different channels. MUP \cite{MUP} is one of
the MAC layer protocols designed to aggregate the bandwidth of multiple radios
tuned to different channels while communicating with a neighbor. GLL, however,
introduces a generic link layer (GLL) approach to use multiple radios while
communicating with a certain destination \cite{GLL}. Such approach is unique
because it considers using an intermediate relay node while communicating with
the destination. GLL assumes that their are two kinds of radios: (1) radios
that are directly connected to the destination, and (2) radios that are
connected to the destination though a one-hop relaying node.

\section{Layer-independent Features}
\label{LIF}
We define the layer-independent features as the set of features shared by
bandwidth aggregation solutions regardless of the layer in which they are
implemented. These features include: (1) interfaces characteristics estimation,
(2) applications characteristics estimation, (3) scheduling, and (4) network
support and communication model.

\subsection{Interface Characteristics Estimation}
Interface characteristics estimation is one of the most important features of
any bandwidth aggregation system. It is responsible for capturing the
heterogeneity of the different network interface including traffic load, loss
rate, interface capacity, etc. In this section, we discuss the interface
characteristics estimation techniques used by various solutions as well as the
different approaches proposed to estimate each of them.

\subsubsection{Bandwidth estimation}
Estimating the available bandwidth at each interface is a key functionality for
bandwidth aggregation systems since it is the most popular metric taken into
account when scheduling data across different interfaces. The most dominant
techniques proposed for such estimation are: (1) delay jitter based estimation,
(2) packet pair, (3) interface traffic monitoring, (4) probing reference
servers, (5) operator-assisted estimation, and (6) implicit estimation.

\begin{figure}[!t]
  \centering
  \includegraphics[width=\columnwidth]{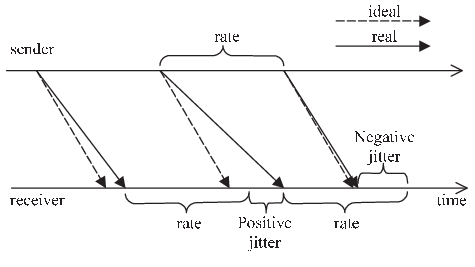}\\
  \caption{Negative and positive inter-arrival time jitter values caused by the 
  variance in propagation delay while average jitter hovers around zero \cite{R-MTP}.}\label{jitter}
\end{figure}

\textit{\textbf{- Delay-Jitter-Based Estimation:}} R-MTP \cite{R-MTP}
implements a delay jitter based bandwidth estimation technique, which is the
receiver is responsible of doing it. This technique is based on an agreement
regarding the transmission data rate between the sender and receiver. The
receiver estimates the delay jitter based on the inter-arrival time between
packets (Figure~\ref{jitter}). The average long term jitter should hover around
zero since it takes positive and negative values as shown in
Figure~\ref{jitter}. In case of congestion, this jitter will increase, and
thus, the receiver will be able to detect congestion and notify the sender that
the available bandwidth is less than the utilized data rate. The receiver also
estimates the reception data rate (available bandwidth) and sends it to the
sender. Using this technique, however, the sender can only reduce its sending
rate but cannot detect the increase of the available bandwidth. This leads to a
waste of the available bandwidth unless detected using other bandwidth
estimation techniques. Overcoming this problem can be done by periodically
probing the paths using other bandwidth estimation techniques or by enabling
the sender to increase the transmission rate in case of being able to maintain
the current transmission rate for certain period of time.

\textit{\textbf{- Packet-Pair:}} The packet-pair technique \cite{SBAM_13} is
one of the popular bandwidth estimation techniques used by several bandwidth
aggregation systems
\cite{OSCAR,OSCAR-MASS,networkTCP,NetworkUDP,deployableSCTP,SBAM}. In addition,
R-MTP \cite{R-MTP} uses it to overcome some of the shortcomings mentioned
above. In this technique, the sender sends two back-to-back packets on each
path. These packets are served by the path bottleneck, which leads to spacing
them out in time. Once a packet arrives at the destination, an ACK is directly
sent back to the source. These ACKs will preserve the same time spacing between
the reception of the packets. By measuring the inter-arrival time between the
ACKs, the available bandwidth at the path bottleneck can be estimated. The
sending rate can then be adjusted based on this estimate.

Note that to gather accurate bandwidth estimates using the packet-pair
technique, the sender should use long packet trains (not only two packets) and
measure the average inter-arrival time between every two consecutive ACKs. To
avoid the overhead of probing the network with long packet trains, the systems
that adopts packet-pair techniques use application data packets to probe the
network.

\textit{\textbf{ - Interface Traffic Monitoring:}} Several approaches rely on
monitoring the different network interfaces to estimate the
bandwidth\cite{DNIS,OPERETTA,DBAS,G-DBAS,DBAS-ComNets}. They estimate the
available bandwidth by measuring the average number of bytes sent and received
per interface when running TCP connections. This is based on the fact that TCP
congestion control enables it to transmit at a data rate close to the available
bandwidth. Although this technique accurately estimate the bandwidth of a
downlink, It is not suitable for estimating the bandwidth of an uplink in case
of having UDP streams uploading traffic concurrently with TCP connections.

\textit{\textbf{- Probing Reference Servers:}} Assuming that the bottleneck is
local, Habak et al. use geographically dispersed reference servers to estimate
the available bandwidth at each interface
\cite{DBAS,DBAS-ComNets,G-DBAS,OPERETTA}. They periodically connect to these
servers to estimate the uplink and the downlink available bandwidth. They also
combine these estimates with statistics collected during the normal data
transfer by interface traffic monitoring. When running in packet-oriented mode,
where each packet can be scheduled to a different interface, DBAS
\cite{DBAS,DBAS-ComNets} and OPERETTA\cite{OPERETTA}, obtain a better estimate
of the available bandwidth by probing the actual destination.

Intentional Networking \cite{Intentional} uses a similar method where bandwidth
estimation is based on randomly probing selected geographically spanned
reference servers \cite{IntenActive}. Basically, they connect to specific ports
on such servers and open a TCP connection where the server uses the TCP
protocol to send data as fast as possible through such connection. Their
mechanism terminates the connection after 1 second and uses the measured data
rates to estimate the available bandwidth. To avoid the effect of the TCP's
slow start, they discard the estimates gathered during the first 500
milliseconds. To enhance these estimates and minimize the probing overhead,
they rely on data packets, if available, to estimate the
available bandwidth between the source and the actual destination
\cite{IntenPassive}.  To achieve this, They monitor the exchanged data packets
between a source and destination, and estimate the available bandwidth
accordingly between them using exponential averaging.  Furthermore, they deploy
four filters on their observations in order to quickly detect network status
changes while resisting transients in these observations.

\textit{\textbf{- Operator-assisted estimation:}} For bandwidth estimation,
MOTA makes the network operator aid the host in estimating the needed bandwidth
from this operator \cite{mota}. The operator broadcasts information about its
available bandwidth and current load. This information is then used by the host
to estimate the bandwidth it will utilize if its load traverses the
corresponding interface. This approach is based on the willingness of network
operators to share accurate information about their available bandwidth and
current load.

\textit{\textbf{- Implicit estimation:}} Other systems depend on their
congestion control mechanisms to keep the transmission rate at each interface
close to its available bandwidth
\cite{GMCC-Coop,LS-SCTP,IPCC-SCTP,pTCP,RCP,cmpTCP,CMTSCTP,cTCP,WiMP-SCTP,cmpSCTP,huang2010design,MPTCP,E2EMPT,wischik2011design,mpsctp2011,cao2012delay,paasch2012exploring,MPSCTP,AMTCP,multipathCubic,BinomialCC,QoSTCP,raiciu2012hard,shamszaman2013feasibility,peng2013multipath,SC-MPTCP,DRePaS}.
Although they adopt different congestion control mechanisms
(Section~\ref{sec:Transport}), they rely on the same concepts to avoid
explicitly estimating the available bandwidth for each interface.
Unfortunately, this technique is only applicable when developing a reliable,
congestion-aware, and multi-interface-aware transport layer protocol which
faces a steep deployment barrier.

\subsubsection{Delay}
Bandwidth aggregation solutions estimate the end-to-end delay to be used in
loss detection and/or scheduling. Different bandwidth aggregation solutions,
however, estimate this delay in different forms based on their main purpose of
estimating it. For instance, many solutions, specially transport layer
protocols,  estimate the round trip time (RTT) as the average time difference
between sending a packet and receiving its acknowledgement and use it to
calculate the retransmission timeout (RTO)
\cite{MuniSocket,ETOM,NetworkUDP,CFPS,FPS,LS-SCTP,IPCC-SCTP,pTCP,RCP,cmpTCP,CMTSCTP,cTCP,WiMP-SCTP,cmpSCTP,huang2010design,MPTCP,E2EMPT,wischik2011design,mpsctp2011,cao2012delay,paasch2012exploring,MPSCTP,GMCC-Coop,multipathCubic,AMTCP,BinomialCC,QoSTCP,raiciu2012hard,shamszaman2013feasibility,peng2013multipath,SC-MPTCP,DRePaS}.
On the other hand, to take the delay into account while distributing traffic
across interfaces, Phatak et al. argue that it is sufficient to calculate the
differences in latency between the interfaces, not the actual latency of each
interface \cite{Novel}. To avoid requiring time synchronization that is
infeasible, all time calculations are measured at the sender. The sender sends
multiple packets at the same time on different interfaces and calculates the
difference between receiving their ACKs. To increase the accuracy of this
calculation, they force the receiver to send all the ACKs using a single path.
Hence, these ACKs will face the same delays and the dominant part of their
reception time difference is due to the difference of the main packets
reception. ATLB also estimates the senders's queuing delay as the ratio between
the queue length of each sub-flow along with its average achieved throughput
\cite{hasegawa2007}. Cross-layer FPS uses a cross layering approach to estimate
the MAC layer contention incurred delays (ex: backoff time) \cite{CFPS}.

\subsubsection{Energy Consumption} With the increased adoption of mobile
battery-operated devices, taking energy consumption into account while building
a bandwidth aggregation system becomes crucial. Hence, estimating energy
consumption rates of the each network interface becomes one of their critical
tasks. Habak et al. rely on the fact that energy consumption is based only on
the interface's NIC\cite{G-DBAS,OPERETTA,OSCAR,OSCAR-MASS}. Hence, they save
the various energy consumption rates of different network cards in a database.
Interface characteristics estimation modules can then query this database to
estimate the interfaces' energy consumption rates. On the other hand, GreenBag
models the energy consumption of wireless interfaces (Wifi and LTE) as a
function of the transmission time and the bandwidth used, as well as other
constant factors that are technology dependent \cite{GreenBag}. Their model
takes into account the energy consumed in the active transmission/reception
states and the TAIL state.

\subsection{Applications Characteristics Estimation} Application
characteristics knowledge can significantly affect the decisions taken by a
bandwidth aggregation solutions. Recent measurement study shows how application
characteristics can affect the efficiency of a bandwidth aggregation solution
(MPTCP \cite{wischik2011design}) \cite{WifiLTEBoth}. Therefore, a number of
solutions use their knowledge about the applications characteristics to enhance
scheduling decisions. Based on the method used to obtain such knowledge,
solutions can be classified into three main categories: (1) qualitative input,
(2) quantitative input, and (3) estimation.

\subsubsection{Qualitative input} In this case, the system takes hints from the
applications to enhance their scheduling technique. For instance, Intentional
Networking asks applications to determine their type (foreground or background)
and their transmission load (small or large) from some predefined categories
\cite{Intentional}. On the other hand, many systems ask applications to
determine their required reliability level in either course-grained granularity
(reliable or not) \cite{LS-SCTP,SCTP,cmpSCTP,W-SCTP,FPS,CFPS,IPCC-SCTP} or
fine-grained granularity through specifying a set of data-reliability
constraints\cite{Intentional}.

\subsubsection{Quantitative input} Some systems ask applications to explicitly
define their required bandwidth or associate traffic with deadlines
\cite{MMTP,QAWBA,ALP-A}. On the other hand, MOTA \cite{mota} takes application
weights from the user to know the importance of each application. Bandwidth is
assigned to applications according to their relative weight.

\subsubsection{Estimation} To increase the ease of deployment, other systems
estimate application requirements instead of explicitly asking applications to
determine what they need. This approach has the advantage of being backwards
compatible and transparent to the applications. For example, Saeed et al.
\cite{DNIS} and Habak et al. \cite{DBAS,DBAS-ComNets,OPERETTA}  estimate the
connection's sent and received bytes based on each application connections'
history. As a result, the estimated connection length equals the average
connection length calculated from the history. Another technique used by these
systems is based on the application name/type. For example, Skype is treated as
a realtime application while an FTP client is treated as a bulk transfer
application. Because application layer protocols generally have reserved ports
for their communication, OSCAR uses a similar approach but instead of
maintaining these estimates for each application, it maintains it for each port
number\cite{OSCAR,OSCAR-MASS}. GreenBag, however, monitors the status of the
video player and estimates whether it is playing or buffering \cite{GreenBag}.
It also estimates the video quality, played time, and remaining time.

\subsection{Scheduling}
\label{Scheduling}
Scheduling data across the available interface is the center piece of any
bandwidth aggregation system. It utilizes all the available information at the
system to take the best scheduling decisions. In this section, we address two
aspects of scheduling: scheduling granularity and scheduling techniques.

\subsubsection{Scheduling granularity} Scheduling granularity refers to the
unit of data that can be assigned to a network interface. There are two
categories for scheduling granularity: packet-level and connection-level.

\textit{\textbf{- Packet-Level Scheduling:}} To achieve optimality, most
bandwidth aggregation solutions adopt packet-level scheduling granularity, in
which packets belonging to the same connection can be assigned to different
interfaces. This requires support from both ends of the connection, or the
introduction of proxy servers, and usually leads to higher performance.

\textit{\textbf{- Connection-Level Scheduling:}} To ease the deployment, some
bandwidth aggregation solutions adopt connection-level scheduling granularity,
in which different connections can be assigned to different network interfaces.
However, packets belonging to the same connection must be assigned to the same
network interface. These solutions either utilize connection-level scheduling
as their main operational mode \cite{DNIS,G-DBAS} or as an optional mode
triggered by the lack of network support \cite{MAR,DBAS,DBAS-ComNets,OPERETTA}.
MOTA  is considered a special case of connection-oriented scheduling in which
connections belonging to the same application are assigned to the same network
interface \cite{mota}. The main advantage of connection-oriented scheduling is
backwards compatibility with legacy servers. In contrast, Habak et al. show
that connection-level scheduling granularity can significantly enhance
performance if and only if connection lengths are taken into account while
scheduling \cite{DBAS}. Otherwise, It can lose its advantage, and in some cases
lead to performance degradation.

\subsubsection{Scheduling techniques} In this section we present the most
prominent scheduling techniques that have been proposed to distribute data
across different interfaces.

\textit{\textbf{- Round Robin:}} Round-robin scheduling is used in MAR
\cite{MAR}, and adopted as a baseline technique for comparison in a number of
systems \cite{DNIS,DBAS,OPERETTA}. This technique assigns data to interfaces in
a rotating basis without taking into account the capacity of the interface or
the application requirements. SRR investigated a queue-size-based variant of
the round-robin scheduler \cite{SRR}. In this variant, the schedule iterates on
the interfaces in a rotating basis assigning each packet to the interface that
has free slots in their queue.

Many researchers investigated using weighted round robin scheduling. Some of
them weighted the scheduling by the available bandwidth of each interface
\cite{DBAS,DBAS-ComNets,G-DBAS,OPERETTA,NetworkUDP,nguyen2013tcp,kim2008packet}.
LS-SCTP, however, defines the weights as the ratio between the congestion
window size and the round trip time, which is considered an estimate of the
available bandwidth \cite{LS-SCTP}.

Another solution is based on a mathematical model to determine the fraction of
packets that should be sent by each interface without degrading TCP
performance\cite{Novel}. The idea is to make all interfaces have the same
timeout value. This is based on scheduling packets over different interfaces
based on their relative bandwidth, similar to the weighted round robin
technique. However, contrary to weighted round robin, which has a fixed packet
size, the proposed solution has a different packet size for each interface to
guarantee the same timeout value on all interfaces.

\textit{\textbf{- Maximum Throughput:}} Assuming we are bound by
connection-level granularity scheduling, Habak et al. introduce a maximum
throughput scheduling technique \cite{DBAS,DBAS-ComNets}. This technique aims
to maximize the overall system bandwidth without considerations to the
bandwidth of a specific connection or stream. It works in the
connection-oriented granularity mode. For a new connection, the scheduler
assigns it to the network interface that will maximize system throughput. This
is equivalent to assigning the new connection to the interface that minimizes
the time needed to finish the current system load in addition to the load
introduced by this new connection. This algorithm depends on the estimated
connection length and the estimated bandwidth for each interface.

If packet-level granularity scheduling is possible, minimizing the packet
delivery time reflects the increase in overall system throughput. This approach
maximizes the stream/connection throughput while minimizing the reordering
overhead. Packet-pair based earliest-delivery-path-first scheduling sends
packets in pairs on the path, which will deliver it in the shortest time to the
destination \cite{networkTCP}. Westwood SCTP (W-SCTP) estimates the chunk's
delivery time at each network interface in order to select the interface that
has the shortest delivery time to serve that chunk \cite{W-SCTP}. This
procedure is done for each chunk until the congestion windows of the available
paths have been exhausted. FPS dynamically estimates the packet delivery time
at each network interface and fills this difference with in-order data to avoid
packet reordering \cite{FPS}. Cross-layer FPS \cite{CFPS} extends FPS
\cite{FPS} using a cross-layering approach to include the MAC layer contention
delays (ex. backoff delay) while estimating the packet delivery time at each
interface. MAC layer solutions also maximize throughput by keeping the spectrum
busy. These solutions assign packets to interfaces that have available free
spectrum/media \cite{Bondingplus,PPPMul,MPPP,GLL,MUP}. ATLB ranks the
interfaces such that the minimum score interface has the minimum delivery time,
and assigns packets to the interface with the minimum score
\cite{hasegawa2007}. ATLB calculates the score of each interface using the
following equation: \[ score_{i} = \frac{Q_i}{G_i} + \frac{RTT_i}{2} \] such
that $Q_i$ is the queue length of sub-flow $i$, $G_i$ is the average throughput
of sub-flow $i$ and $RTT_i$ is its average round trip time.

\textit{\textbf{- Bandwidth Delay Product:}} SBAM schedules data based on the
bandwidth delay product (BDP) of each network interface \cite{SBAM}. The
technique starts by sending data on the interface with the maximum BDP. If the
other end supports SBAM, the system leverages the other interfaces and packets
are distributed over the different interfaces according to their BDP.

\begin{figure}[!t]
  \centering
  \includegraphics[width=\columnwidth]{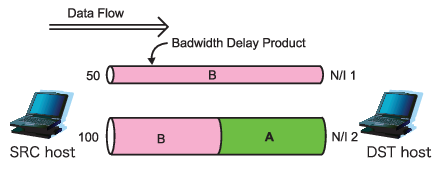}\\
  \caption{SBAM data distribution relative to the interfaces bandwidth-delay product (BDP) \cite{SBAM}.}\label{DataTransmission}
\end{figure}

\textit{\textbf{- Leveraging Congestion Control:}} Distributing packets across
different interfaces can also be done based on the congestion control mechanism
\cite{LS-SCTP,IPCC-SCTP,pTCP,RCP,cmpTCP,CMTSCTP,cTCP,WiMP-SCTP,cmpSCTP,huang2010design,MPTCP,AMTCP,E2EMPT,GMCC-Coop,wischik2011design,mpsctp2011,cao2012delay,paasch2012exploring,MPSCTP,multipathCubic,BinomialCC,QoSTCP,SC-MPTCP,raiciu2012hard,shamszaman2013feasibility,peng2013multipath,DRePaS}.
Although systems adopting this approach implement different congestion control
mechanisms (Section~\ref{sec:Transport}), they share the concept of using these
mechanisms for scheduling. In this case they apply congestion control
mechanisms on each interface, or on the whole connection while dividing the
congestion window over the interfaces and assigning packets to interfaces only
when they have empty space in their congestion window. This technique maintains
the transmission rate on each interface close to its full capacity to increase
the overall system throughput. Furthermore, E2EMPT uses a path priority
assignment scheme to be used while assigning packets to interfaces in case of
having multiple interfaces with free slots in their congestion windows
\cite{E2EMPT}.

\textit{\textbf{- Rate-based:}} Magalhaes et al. use rate-based techniques to
schedule packets \cite{R-MTP,MOPED}. After estimating the available bandwidth
for all network interfaces, this technique calculates the packet rate that each
interface can support. It sends packets on each interface with the rate
supported by this interface. Such technique relies on accurately estimating the
capacity of each interface.

\textit{\textbf{- Energy and Cost Aware:}} G-DBAS proposes two types of
energy-aware scheduling techniques: an energy efficient scheduler and a
utility-based scheduler \cite{G-DBAS}. The energy-efficient scheduler assigns
connections to the interface that minimizes the overall energy consumption. The
utility-based scheduler, however, combines the maximum throughput and energy
efficient schedulers using a user defined utility function to achieve different
user goals.

OPERETTA combines energy efficiency with throughput maximization by formulating
the scheduling problem as a linear program\cite{OPERETTA}. Its main target is
to minimize the energy consumption while achieving a certain amount of
throughput. This throughput is calculated from a user defined utility function
that indicates the user's willingness to increase energy consumption for more
throughput. OPERETTA also combines the packet-oriented mode with the
connection-oriented mode reaching the optimal target without updating any all
destination servers. On the other hand, OSCAR combines energy efficiency and
cost efficiency with throughput maximization by proposing multi-objective and
multi-modal scheduling \cite{OSCAR,OSCAR-MASS}. Based on the operational mode,
OSCAR tries to optimize one parameter while maintaining the other parameters
within the user-accepted ranges. It also combines the packet-oriented mode with
the connection-oriented mode reaching the optimal target without updating all
destination servers.

GreenBag schedules the real-time video streaming traffic across interfaces to
support the required quality of service in the most energy efficient way
\cite{GreenBag}. It divides the video file to chunks and downloads each chunk
on the interface that minimizes both playback time and energy consumption. On
the other hand, ALP-A assigns HTTP requests to the minimum energy consuming
interface as long as it satisfies the application defined QoE requirements
\cite{ALP-A}.

\begin{figure}[!t]
  \centering
  \includegraphics[width = \columnwidth]{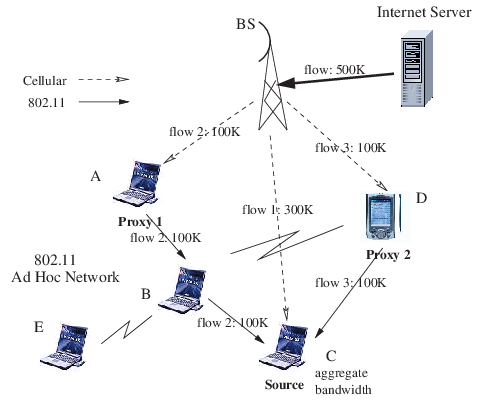}\\
  \caption{QAWBA collaborative bandwidth aggregation form example \cite{QAWBA}.}\label{QAWBA}
\end{figure}

\textit{\textbf{- Quality of Service:}}
Another approach is application-required QoS based packet scheduling. For
instance, in QAWBA, when aiming for collaborative bandwidth aggregation between
peers \cite{QAWBA}, each node is assumed to have one interface connected to the
Internet and another connected with its peers. The application first defines
its required bandwidth. Then the scheduler reserves as much as possible from
its local link and sends requests to other nodes to reserve the extra bandwidth
needed along with the maximum number of hops for this request. Scheduling
packets is performed to fit the reserved bandwidth at each path. Figure
\ref{QAWBA} shows an example of QAWBA where five mobile nodes form a MANET. The
client node $C$ runs an application requiring 500Kbps, of which only 300Kbps
are available from its cellular link. $A$ and $D$ act as proxies to forward a
portion of the total traffic to $C$. The 500Kbps traffic flow is split into
three flows at the base station, and then forwarded to $C$ via different paths.
Thus, with the help of nodes $A$ and $D$, C is able to receive the required
500Kbps bandwidth by aggregating three flows, which would not be possible under
one single cellular connection.

MMTP provides another scheduling technique designed for multimedia applications
that have hard deadlines in delivering frames \cite{MMTP}. This technique
selects the best interface to send a frame based on its estimated arrival time
using this interface and its arrival time deadline. For a given frame, it
searches for network interfaces that have non-utilized available bandwidth.
This is achieved in the protocol by making each interface generate tokens
according to its own available bandwidth. The tokens are used based on the
following rules: 
\begin{itemize}
	\item If no token is available, the frame must wait.
        \item If exactly one token is available, and the estimated arrival time
            is before the frame arrival deadline, then send the frame using
            this interface. Otherwise, wait until either a new token appears to
            check for its interface, or the deadline is reached after which it
            will be dropped.
        \item If more than one token is available, select the interface which
            has the highest propagation delay that can deliver the packet on
            time. This keeps the interface channel filled and increases the
            probability that a new packet will be delivered on time.
\end{itemize}

\subsection{Network Support and Communication Model}
\begin{figure}[!t]
  \includegraphics[width=\columnwidth]{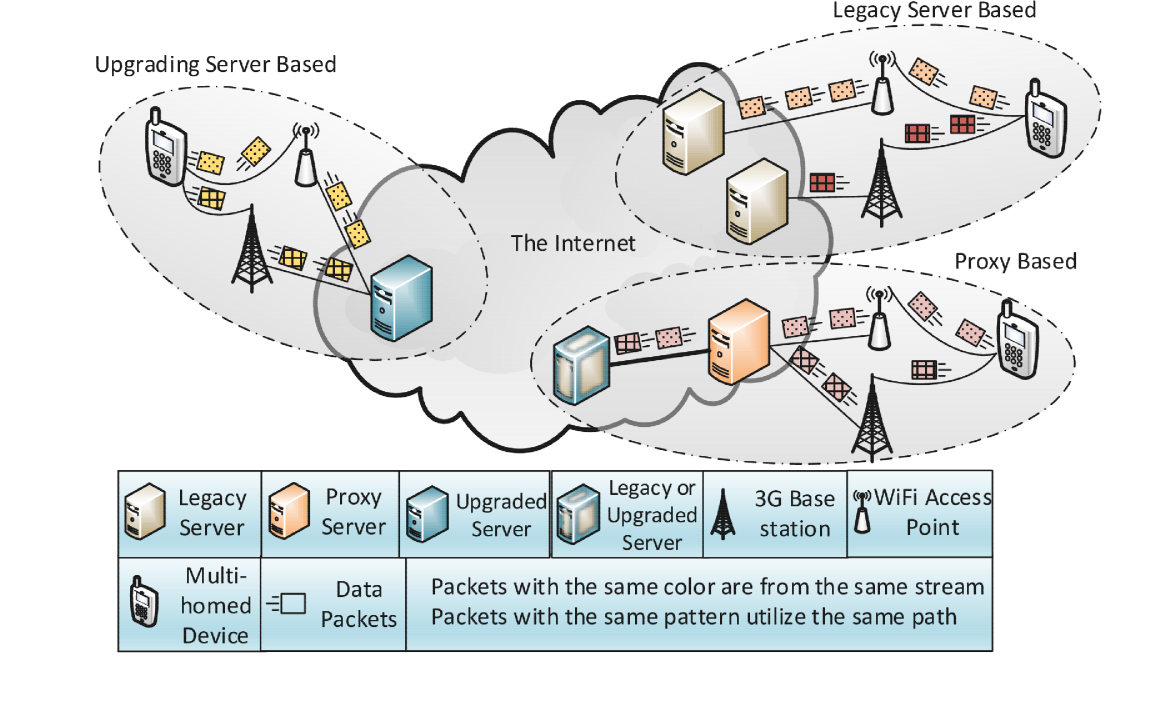}\\
  \caption{Bandwidth aggregation systems adopted client-server communication models and their required network support.}\label{SCNFigure}
\end{figure}

Different systems adopt different client-server communication models, so they
require different levels of network support. These models are selected to
achieve certain system objectives such as increasing the overall system
performance and minimizing the system adoption cost. The proposed systems
generally adopt three communication models that we visually summarize in
Figure~\ref{SCNFigure}: 1) upgraded-server-based communication, 2) proxy-based
communication  and 3) legacy-server-based communication.

\subsubsection{Upgraded-Server-based Communication Model} This communication
model mainly targets increasing the overall system performance without
requiring any updates to the core network infrastructure. Therefore, it relies
on implementing bandwidth aggregation solutions at the communication end-points
(the client and the server). Hence, many bandwidth aggregation systems adopt
this model either in their normal operation mode
\cite{PPPMul,Intentional,SRR,BinomialCC} or as an optional mode for improving
the system's performance \cite{DBAS,OPERETTA,DBAS-ComNets}. With the
flexibility of upgrading communicating end points in this model, the ability to
design fine-grained schedulers and accurate interface characteristics
estimators increases. On the other hand, this approach carries the overhead of
upgrading destination servers in order for it to be adopted.

\subsubsection{Proxy-based Communication Model} Another communication model is
to avoid or minimize the overhead of upgrading servers by placing a proxy
server that is aware of the client's multiple interfaces and aids the client in
its bandwidth aggregation. In such cases, the proxy is assumed to be connected
to destination servers via high bandwidth links. Most solutions adopting this
communication model mainly use it to avoid upgrading end servers
\cite{deployableSCTP,TMCMCC}. PRISM \cite{PRISM} deploys this proxy server to
minimize the amount of new functionalities that should be supported by the
upgraded communication server.

MAR's optional mode of operation is a special case of adopting this model
\cite{MAR}. The main difference is that they do not assume that clients are
equipped with multiple network interfaces, but instead, are connected to a MAR
router equipped with multiple network interface. The MAR router works with the
proxy server in this optional mode in order to utilize its available network
interfaces and enhance the overall system performance.

Overall, this model of communication enables clients to use fine-grained
scheduling techniques as well as efficient interface characteristics estimation
solutions. On the other hand, a proxy introduces another set of challenges such
as where to place it and how to avoid multiple clients contending for the proxy
server, which can ultimately render the proxy server itself becoming a
bottleneck.

\subsubsection{Legacy-Server-based Communication Model} Driven by the need to
avoid infrastructure and server updates, some work has adopted this
communication model. Although this model lacks the ability to deploy
fine-grained scheduling techniques, its coarse-grained (connection-oriented)
scheduling enables clients to utilize available interfaces and achieve high
performance gains. Researchers adopted this communication model while
developing bandwidth aggregation systems that put the responsibility of
exploiting interfaces on the client devices which has the minimum updating cost
\cite{OPERETTA,mota}.

\section{Evolution and Challenges}
\label{Evol}
After providing an overview of bandwidth aggregation systems and their features, this section discusses the evolution of these systems along with current challenges that remain to be addressed by the research community.

\subsection{Evolution}
Bandwidth aggregation solutions have evolved over the past decade in different forms. We chronologically present the most prominent research conducted in this area in Table~\ref{Evol-Table}, and compare them based on the set of parameters shown in the table. In addition, we discuss in this section two forms of evolution: Protocol stack layered evolution and scheduling granularity evolution.

\begin{table*}[!t]
\centering
\small
\setlength{\extrarowheight}{0.5pt}
\begin{tabular}[t]{!{\VRule[1.5pt]}c!{\VRule[1.5pt]}c!{\VRule[0.5pt]}c!{\VRule[1.5pt]}c!{\VRule[0.5pt]}c!{\VRule[0.5pt]}c!{\VRule[0.5pt]}c!{\VRule[1.5pt]}c!{\VRule[0.5pt]}c!{\VRule[0.5pt]}c!{\VRule[1.5pt]}c!{\VRule[0.5pt]}c!{\VRule[1.5pt]}c!{\VRule[1.5pt]}}\specialrule{1.5pt}{0pt}{0pt}

 & \multicolumn{2}{c!{\VRule[1.5pt]}}{Sched. Gran.} & \multicolumn{4}{c!{\VRule[1.5pt]}}{Network Stack Layer} & \multicolumn{3}{c!{\VRule[1.5pt]}}{Deployability} & \multicolumn{2}{c!{\VRule[1.5pt]}}{App. Ch.} &\\\cline{2-10}

System & Packet & Con.  & App. & Tran. & Net. & MAC & Client & Server & Infra.  & \multicolumn{2}{c!{\VRule[1.5pt]}}{Knowledge} & Year\\\cline{11-12}

 & Level & Level & Layer & Layer & Layer & Layer & Upd. & Upd. & Supp. & Input & Est &  \\\specialrule{1.5pt}{0pt}{0pt}

 MP \cite{PPPMul} & \checkmark  &  &  &  &  & \checkmark & \checkmark & \checkmark &  & & & 1996 \\[0.5pt]\hdashline[2pt/0pt] 
 
 MMP \cite{MPPP}  & \checkmark   &  &  &  &  & \checkmark & \checkmark & \checkmark &  & & & 1999 \\[0.5pt]\hdashline[2pt/5pt]
 
 SRR \cite{SRR}  & \checkmark   &  &  &  & \checkmark &  & \checkmark &  \checkmark &  & & & 1999 \\[0.5pt]\hdashline[2pt/0pt] 
 
 MMTP \cite{MMTP}  & \checkmark   &  &  & \checkmark &  &  & \checkmark & \checkmark & &  \checkmark & & 2001 \\[0.5pt]\hdashline[2pt/5pt]
 
 R-MTP \cite{R-MTP}  & \checkmark   &  &  & \checkmark &  &  & \checkmark & \checkmark &  & & & 2001 \\[0.5pt]\hdashline[2pt/0pt]
 
 Phatak et al. \cite{Novel}  & \checkmark &    &  &  & \checkmark &  & \checkmark & \checkmark &  &  & & 2002 \\[0.5pt]\hdashline[2pt/5pt] 
 
 MuniSocket \cite{MuniSocket}  & \checkmark   &  & \checkmark &  &  &  & \checkmark & \checkmark &  &  & & 2002\\[0.5pt]\hdashline[2pt/0pt]
  
 SCTP \cite{SCTP}  & \checkmark &  &  &   \checkmark &  &  & \checkmark & \checkmark & &  \checkmark & & 2003 \\[0.5pt]\hdashline[2pt/0pt]
 
 BondingPlus\cite{Bondingplus}  & \checkmark &  &  &    &  & \checkmark & \checkmark & \checkmark & & & & 2004 \\[0.5pt]\hdashline[2pt/5pt]
 
 MUP \cite{MUP}  & \checkmark &  &  &    &  & \checkmark & \checkmark & \checkmark & & & & 2004 \\[0.5pt]\hdashline[2pt/5pt]
 
 MAR \cite{MAR}  & \maltese & \checkmark   &  &  & \checkmark &  &  &  & \checkmark  & & & 2004 \\[0.5pt]\hdashline[2pt/5pt]
  
 QAWBA \cite{QAWBA} & \checkmark   &  &  &  & \checkmark &  & \checkmark &  & & & & 2004 \\[0.5pt]\hdashline[2pt/5pt] 
 
 LS-SCTP \cite{LS-SCTP}  & \checkmark  &  &  & \checkmark &  &  & \checkmark & \checkmark &  &\checkmark & & 2004 \\[0.5pt]
 
 IPCC-SCTP \cite{IPCC-SCTP}  & \checkmark  &  &  & \checkmark &  &  & \checkmark & \checkmark &  &\checkmark & & 2004 \\[0.5pt]
  
 W-SCTP \cite{W-SCTP}  & \checkmark &  &  &   \checkmark &  &  & \checkmark & \checkmark & & \checkmark & & 2004 \\[0.5pt]\hdashline[2pt/5pt]
 
 mTCP \cite{mTCP}  & \checkmark &  &  &   \checkmark &  &  & \checkmark & \checkmark & & & & 2004 \\[0.5pt]\hdashline[2pt/0pt]
 
 
 GLL \cite{GLL}  & \checkmark &  &  &  &  &   \checkmark & \checkmark & \checkmark &  & & & 2005 \\[0.5pt]\hdashline[2pt/5pt] 
  
 Chebrolu et al. \cite{networkTCP}  & \checkmark   &  &  &  & \checkmark &  & \checkmark &  &   \checkmark & & & 2005 \\[0.5pt]\hdashline[2pt/5pt]
  
 pTCP \cite{pTCP}  & \checkmark &    &  & \checkmark &  &  & \checkmark & \checkmark & &  & & 2005 \\[0.5pt]
 
 $R^2$CP \cite{RCP}  & \checkmark &  &    & \checkmark &  &  & \checkmark & \checkmark &   & & & 2005 \\[0.5pt]\hdashline[2pt/5pt] 
 
  
 PRISM \cite{PRISM} & \checkmark &  & \maltese & \checkmark & \maltese &  & \checkmark & \checkmark  & \checkmark & & & 2005 \\[0.5pt]\hdashline[2pt/0pt]
 
 CMT \cite{CMTSCTP}  & \checkmark &   &  & \checkmark &  &  & \checkmark & \checkmark & & \checkmark & &  2006 \\[0.5pt]\hdashline[2pt/5pt]
 
 cmpTCP \cite{cmpTCP}  & \checkmark &   &  & \checkmark &  &  & \checkmark & \checkmark &  & & & 2006 \\[0.5pt]\hdashline[2pt/5pt] 
 
  SBAM \cite{SBAM}  & \checkmark &  &  \checkmark &  &  &  & \checkmark & \checkmark & & & &  2006 \\[0.5pt]\hdashline[2pt/0pt]
  
  
 WiMP-SCTP \cite{WiMP-SCTP}  & \checkmark &  &  &   \checkmark &  &  & \checkmark & \checkmark &  & \checkmark& & 2007 \\[0.5pt]\hdashline[2pt/5pt]
  
 cTCP \cite{cTCP}  & \checkmark &  &   & \checkmark &  &  & \checkmark & \checkmark & & & &  2007 \\[0.5pt]\hdashline[2pt/5pt]
 
 ATLB \cite{hasegawa2007}  & \checkmark &  &   & \checkmark &  &  & &  & \checkmark  & &  & 2007  \\[0.5pt]\hdashline[2pt/5pt] 
  
 Handheld R. \cite{TMCMCC} & \checkmark   &  & \checkmark &  &  &  & \checkmark &  & \checkmark & & &  2007 \\[0.5pt]\hdashline[2pt/5pt]
  
 COMBINE \cite{combine} & \checkmark  &  & \checkmark &  &  &  & \checkmark & \checkmark &  & & &  2007 \\[0.5pt]\hdashline[2pt/0pt] 
 
 
 cmpRTCP \cite{cmpRTCP}  & \checkmark &   &  & \checkmark &  &  & \checkmark & \checkmark & & \checkmark & & 2008 \\[0.5pt]
 
 cmpSCTP \cite{cmpSCTP}  & \checkmark &  &  &   \checkmark &  &  & \checkmark & \checkmark &  & \checkmark& &  2008 \\[0.5pt]\hdashline[2pt/0pt]
 
 Evensen et al. \cite{NetworkUDP}   & \checkmark   &  &  &  & \checkmark &  & \checkmark &  & \checkmark & & &  2009 \\[0.5pt]\hdashline[2pt/0pt]
 
   
 MPTCP \cite{MPTCP}  & \checkmark &  &  &   \checkmark &  &  & \checkmark & \checkmark & &  & &  2010 \\[0.5pt]
 
 E2EMPT \cite{E2EMPT}  & \checkmark &  &  &   \checkmark &  &  & \checkmark & \checkmark & &  & &  2010 \\[0.5pt]\hdashline[2pt/5pt]
   
 MCMT \cite{huang2010design}  & \checkmark &  &  &   \checkmark &  &  & \checkmark & \checkmark &  & \checkmark& &  2010 \\[0.5pt]
 
 FPS \cite{FPS}  & \checkmark &  &   & \checkmark &  &  & \checkmark & \checkmark  & & \checkmark& &  2010 \\[0.5pt]\hdashline[2pt/5pt] 

 Intent. Net. \cite{Intentional}  & \checkmark   &  & \checkmark &  &  &  & \checkmark & \checkmark &  & \checkmark & & 2010  \\[0.5pt]\hdashline[2pt/5pt]
 
 Kasper et al. \cite{UsingHTTP}   & \checkmark &    & \checkmark &  &  &  & \checkmark &  &  & & &  2010 \\[0.5pt]
 
 Kasper et al. \cite{EnhanceVideo}   & \checkmark   &  & \checkmark &  &  &  & \checkmark &   &  & & &  2010 \\[0.5pt]

 Evensen et al. \cite{QualityAdaptive2}  & \checkmark &  & \checkmark &  &  &  & \checkmark &  &  & & &  2010 \\[0.5pt]\hdashline[2pt/5pt] 
 
 DNIS \cite{DNIS}  &  & \checkmark   & \checkmark &  &  &  & \checkmark &  & & &  \checkmark &  2010 \\[0.5pt]\hdashline[2pt/0pt]
 
  Evensen et al. \cite{MultipathUDPBandwidth}  & \checkmark   &  &  &  & \checkmark &  & \checkmark & &  \checkmark & & & 2011 \\[0.5pt]\hdashline[2pt/5pt] 
 
 MOTA \cite{mota}  &  &   \checkmark &  &  & \checkmark &  & \checkmark &  & \checkmark   & \checkmark & &  2011 \\[0.5pt]\hdashline[2pt/5pt] 
  
 Wischik et al. \cite{wischik2011design}  & \checkmark &  &  &   \checkmark &  &  & \checkmark & \checkmark & &  & &  2011 \\[0.5pt]\hdashline[2pt/5pt]
 
 MPSCTP \cite{mpsctp2011}  & \checkmark &  &  &   \checkmark &  &  & \checkmark & \checkmark &  & \checkmark& & 2011  \\[0.5pt]
  
 X-Layer FPS \cite{CFPS}  & \checkmark &  &    & \checkmark &  &  & \checkmark & \checkmark & & \checkmark & &  2011 \\[0.5pt]\hdashline[2pt/5pt]
 
 Evensen et al. \cite{QualityAdaptive}  & \checkmark &  & \checkmark &  &  &  & \checkmark &  &  & & & 2011 \\[0.5pt]\hdashline[2pt/0pt] 

 ETOM \cite{ETOM}  & \checkmark & & & & \checkmark & & & & \checkmark & & &  2012 \\[0.5pt]\hdashline[2pt/5pt]
 
 Tachibana et al. \cite{deployableSCTP}  & \checkmark &  &  &   \checkmark &  &  & \checkmark &  & \checkmark & & &  2012 \\[0.5pt]\specialrule{1.5pt}{0pt}{0pt}
 
 \multicolumn{13}{c}{}
\end{tabular}
\caption{Evolution ($\checkmark$ means required and $\maltese$  means optional)}
\label{Evol-Table}
\end{table*}

\setcounter{table}{1}


\begin{table*}[!t]
\centering
\small
\captionsetup[longtable]{skip=5pt}
\setlength{\extrarowheight}{0.5pt}
\begin{tabular}[t]{!{\VRule[1.5pt]}c!{\VRule[1.5pt]}c!{\VRule[0.5pt]}c!{\VRule[1.5pt]}c!{\VRule[0.5pt]}c!{\VRule[0.5pt]}c!{\VRule[0.5pt]}c!{\VRule[1.5pt]}c!{\VRule[0.5pt]}c!{\VRule[0.5pt]}c!{\VRule[1.5pt]}c!{\VRule[0.5pt]}c!{\VRule[1.5pt]}c!{\VRule[1.5pt]}}\specialrule{1.5pt}{0pt}{0pt}

 & \multicolumn{2}{c!{\VRule[1.5pt]}}{Sched. Gran.} & \multicolumn{4}{c!{\VRule[1.5pt]}}{Network Stack Layer} & \multicolumn{3}{c!{\VRule[1.5pt]}}{Deployability} & \multicolumn{2}{c!{\VRule[1.5pt]}}{App. Ch.} &\\\cline{2-10}

 System & Packet & Con.  & App. & Tran. & Net. & MAC & Client & Server & Infra.  & \multicolumn{2}{c!{\VRule[1.5pt]}}{Knowledge} & Year\\\cline{11-12}

  & Level & Level & Layer & Layer & Layer & Layer & Upd. & Upd. & Supp. & Input & Est & \\\specialrule{1.5pt}{0pt}{0pt} 
  
 MPSCTP \cite{raiciu2012hard}  & \checkmark &  &  &   \checkmark &  &  & \checkmark & \checkmark &  & \checkmark& &  2012 \\[0.5pt]

 Diop et al. \cite{QoSTCP}  & \checkmark &  &  &   \checkmark &  &  & \checkmark & \checkmark & & \checkmark & &  2012 \\[0.5pt]\hdashline[2pt/5pt] 
 
 Kim et al. \cite{ImpoMPTCP}  & \checkmark &  &  &   \checkmark &  &  & \checkmark & \checkmark & & & &  2012 \\[0.5pt]

 MPCubic \cite{multipathCubic}  & \checkmark &  &  &   \checkmark &  &  & \checkmark & \checkmark & &  & &  2012 \\[0.5pt]

 Le et al. \cite{BinomialCC}  & \checkmark &  &  &   \checkmark &  &  & \checkmark & \checkmark & &  & &  2012 \\[0.5pt]

 Cao et al. \cite{cao2012delay}  & \checkmark &  &  &   \checkmark &  &  & \checkmark & \checkmark & &  & &  2012\\[0.5pt]\hdashline[2pt/5pt]
 
 Miyazaki et al. \cite{miyazaki2012evaluation}  &  & \checkmark  & \checkmark &  &  &  & \checkmark & \checkmark  & & & & 2012  \\[0.5pt]\hdashline[2pt/5pt]

 G-DBAS \cite{G-DBAS}  &  & \checkmark &   \checkmark &  &  &  & \checkmark &  & & &  \checkmark & 2012  \\[0.5pt]\hdashline[2pt/5pt]
 
 DBAS \cite{DBAS}  & \maltese & \checkmark  & \checkmark &  &  &  & \checkmark & \maltese  & & & \checkmark &  2012 \\[0.5pt]

 OPERETTA \cite{OPERETTA}  & \maltese & \checkmark &   \checkmark &  &  &  & \checkmark & \maltese & &  & \checkmark &  2012 \\[0.5pt]\hdashline[2pt/0pt] 

 Nguyen et al. \cite{nguyen2013tcp}  & \checkmark &    &  &  & \checkmark &  & \checkmark &  & \checkmark & & &  2013 \\[0.5pt]\hdashline[2pt/5pt]
 
 MPTCP \cite{peng2013multipath}  & \checkmark &  &  &   \checkmark &  &  & \checkmark & \checkmark & &  & &  2013 \\[0.5pt]
 
 MPTCP \cite{shamszaman2013feasibility}  & \checkmark &  &  &   \checkmark &  &  & \checkmark & \checkmark & &  & & 2013  \\[0.5pt]\hdashline[2pt/5pt] 
 
 GreenBag \cite{GreenBag}  & \checkmark &  &   \checkmark &  &  &  & \checkmark &  & &  & \checkmark  &  2013 \\[0.5pt]\hdashline[2pt/5pt] 
 
 DBAS \cite{DBAS-ComNets}  & \maltese & \checkmark  & \checkmark &  &  &  & \checkmark & \maltese  & & & \checkmark & 2013 \\[0.5pt] \hdashline[2pt/0pt]
 
 DRePaS \cite{DRePaS}  & \checkmark &  &  &   \checkmark &  &  & \checkmark & \checkmark & &  & & 2014  \\[0.5pt]
  
 SC-MPTCP \cite{SC-MPTCP}  & \checkmark &  &  &   \checkmark &  &  & \checkmark & \checkmark & &  & & 2014  \\[0.5pt]\hdashline[2pt/5pt] 
 
 OSCAR \cite{OSCAR}  & \maltese & \checkmark &  &  & \checkmark &  & \checkmark & \maltese & &  & \checkmark & 2014 \\[0.5pt]\hdashline[2pt/0pt]

 AMTCP \cite{AMTCP} & \checkmark &  &  &   \checkmark &  &  & \checkmark & \checkmark &  & & & 2015  \\[0.5pt] 

 GMCC-Coop \cite{GMCC-Coop}  & \checkmark &  &  &   \checkmark &  &  & \checkmark & \checkmark &  & & & 2015  \\[0.5pt] \hdashline[2pt/5pt]
 
 OSCAR \cite{OSCAR-MASS}  & \maltese & \checkmark &  &  & \checkmark &  & \checkmark & \maltese & &  & \checkmark & 2015 \\[0.5pt] \hdashline[2pt/5pt]
 
 ALP-A \cite{ALP-A}  & \checkmark &  & \checkmark &  &  &  & \checkmark &  & & \checkmark & & 2015 \\[0.5pt] \specialrule{1.5pt}{0pt}{0pt}
 
\multicolumn{12}{c}{}
\end{tabular}
\caption{Evolution ($\checkmark$ means required and $\maltese$  means optional) (Continue)}
\label{Evol-TableC}
\end{table*}

\subsubsection{Protocol Stack Layered Evolution}
In the beginning, the need for increasing available bandwidth coupled with the ability to be equipped with multiple network interfaces was only the case in data centers. Researchers proposed increasing the bandwidth of database servers in these centers by connecting them with multiple identical wired cables. Hence, they started implementing bandwidth aggregation solutions at the MAC layer in order to make use of these multiple links \cite{PPPMul}.

With the exponential growth in technology and decreasing cost of electronics, the number of multi-homed devices exponentially increased. These devices include normal desktops, laptops, tablets and smart phones. Researchers started developing solutions for utilizing the available interfaces on such devices. Hence, this problem started to look like an end-to-end network problem rather than a single-hop problem. This is why MAC layer solutions could not be adopted in such cases.

Researchers began to define this problem as a transport layer problem that requires the modification or replacement of current transport layer protocols with multipath aware protocols. They started by proposing new multi-path transport layer protocols \cite{MMTP} and utilizing the multihoming support in existing transport layer protocols like SCTP \cite{SCTP}.  Although many solutions have been proposed, well-known single path transport protocols (e.g. TCP) were already heavily adopted and deployed. Therefore, researchers started proposing modifications to TCP to utilize the available interfaces \cite{pTCP}. Because of the formidable deployment barrier these approaches faced, researchers had to find a work-around to overcome this problem through network or application layer approaches.

To overcome the transport layer solutions deployment barrier, researchers developed network layer solutions that can utilize the network interfaces while hiding them from transport layer protocols to avoid performance degradation \cite{Novel}. Others developed application layer solutions that utilize multiple transport protocol sessions in order to exploit available interfaces \cite{MuniSocket}. The main drawback of such solutions is the need for upgrading the legacy servers or updating the network infrastructures. This drawback encouraged researchers to focus on developing application layer solutions on stand alone devices. Therefore more recent application layer solutions that do not require upgrading end-servers or modifying network infrastructure appeared \cite{DBAS}.

\subsubsection{Scheduling Granularity Evolution}
Once researchers started to think about providing bandwidth aggregation solutions, they were aiming for the optimal performance. Hence, packet level scheduling was initially adopted. All solutions adopted this fine-grained level of scheduling to maximize performance gains. This level of scheduling, however, introduced the challenge of high deployment cost as a result of the need for upgrading end-servers, modifying network infrastructure, and/or updating client applications. This fact is largely the reason we do not see any pervasive bandwidth aggregation solutions to date.

Researchers had to think about this problem differently. Rodriguez et al. \cite{MAR} noticed that applications tend to have many connections that can be scheduled across available interfaces at connection-level granularity to enhance throughput without modifying end-servers. Their main drawback is that they did not design a new connection-based scheduling technique, but used simple round-robin and weighted round-robin techniques which do not guarantee performance enhancement in the long run.

At this stage, some researchers focused on minimizing the cost of using packet level scheduling techniques, while others developed more novel connection-level scheduling techniques \cite{DNIS,DBAS,OPERETTA}. For example, OPERETTA \cite{OPERETTA} combines both connection level scheduling with packet level scheduling to achieve the maximum performance without updating all legacy servers.

\subsection{Challenges}
\label{challenges}
In this section, we discuss some open research challenges which we believe can make use of further investigation by researchers in the community.

\subsubsection{Deployability}
Deploying a bandwidth aggregation system without modification to current infrastructure and devices is one of the most important challenges that prevent most proposed solutions from achieving their ultimate goal. This challenge was not addressed during the past decade until recent work has taken the first steps in doing so \cite{DNIS,DBAS,hasegawa2007}. In this section we address the deployability obstacles that face current solutions and how some solutions have addressed these obstacles. There are four main barriers hindering deployability at different levels: 1) using intermediate devices, 2) upgrading clients, 3) upgrading servers, and 4) modifying applications. After addressing these deployment barriers we shed the light on some attempts to deploy bandwidth aggregation systems.

\textit{\textbf{- Using Intermediate Devices}}
Using intermediate devices to implement bandwidth aggregation solutions limits the adoption of such solution. Requiring such devices, e.g. routers and proxy servers, increases deployment cost. In addition, with the widespread adoption of such solutions, these devices would be performance bottlenecks at edge networks and would need to scale accordingly.

MAR uses a router as well as an optional proxy to implement their solution \cite{MAR}. The router is the device to which their multiple interface clients are attached. Clients connect to this router and set it as their DNS server. The optional proxy's existence changes MAR's mode of operation from connection level scheduling to packet level scheduling.

MOTA \cite{mota} requires updating the network operator's base stations in order to support devices with the information required to schedule the data across different interfaces. It requires the operator to support these devices with information about its current load, the cumulative weight of all the applications assigned to it, and the available bandwidth. Hence, the device uses this information to estimate its available bandwidth.

Other solutions rely on the existence of proxy servers to implement their techniques \cite{hasegawa2007,networkTCP}. In these solutions, client devices connect to the proxy, with the scheduling technique running between them.

On the other hand, other solutions avoid using intermediate devices and implement their solutions only on communicating end-points to minimize their deployment cost \cite{QoSTCP,SBAM}. These solution are more deployable in this sense.

\textit{\textbf{- Upgrading Servers:}}
Upgrading servers restricts the widespread deployment of a solution. It is difficult to upgrade legacy servers around the world to make use of multiple interfaces at the client. Upgrading these servers requires large amounts of money, effort, and time.

Some solutions lost deployability by relying on upgrading servers for their adoption \cite{CMTSCTP,MPTCP}. These solutions focus on building an end-to-end system upgrading both clients and servers to utilize the available interfaces to their maximum. To increase deployability, other solutions avoid upgrading the servers \cite{DNIS,G-DBAS}. Habak et al. \cite{OPERETTA,OSCAR,OSCAR-MASS}, however, have two modes of operation. Connection-oriented mode in which they do not require any updates to the servers in order to increase deployability and a packet-oriented mode which makes use of optional upgrading of servers in order to maximize the overall system performance.

\textit{\textbf{- Modifying Application:}}
Modifying legacy applications greatly impacts bandwidth aggregation system deployment. Generally, bandwidth aggregation solutions implemented below the transport layer do not interface with applications \cite{Novel,networkTCP}. Hence, they are backwards-compatible with current applications. On the other hand, transport layer solutions and application layer solutions differ from one another based on the degree of backward compatibility they offer.

Transport layer protocols that change the contract or interface between the applications and the transport layer are generally not backwards compatible and require modifying legacy applications \cite{MMTP}. Other solutions aim at replacing currently used protocols with new multi-interface aware protocols or extending current protocols to add this awareness \cite{pTCP}. Although such solutions maintain the same interface and contract terms between the applications and transport layer, they require modifying the applications since the only feasible way of deploying them in the current operating system is as a new transport protocol that lies beside the old set of protocols to maintain backwards-compatibility with Internet servers. cTCP \cite{cTCP} solves this issue by implementing a TCP extension that is backwards-compatible with the current TCP by maintaining its interface with the application. Such extension can be deployed in the current operating system as a replacement of the TCP protocol while avoiding the need for updating applications.

On the other hand, different approaches for implementing application-layer solutions exist (Section~\ref{sec:LDFAPP}). Hidden-middlware based solutions do not require updating the applications since they maintain the same interfaces between the applications and their transport layer protocols \cite{DNIS,SBAM}. Non-transparent-middleware based solutions, however, generally require updating the applications because such middleware either updates the interface between the applications and transport layer or requires certain input from these applications \cite{MuniSocket,Intentional}.

\textit{\textbf{- Upgrading Clients:}}
Upgrading clients also impacts deployability. This effect is minimal, however, compared to the other updates above, since it is normal to update devices with new patches nowadays. The impact of this factor depends on the complexity of the updates themselves.

Upgrading the kernel is considered the most complex and expensive update on the client side. Some solutions require updating the client kernel by modifying the network protocol stack  \cite{MPTCP,RCP}. These updates decrease the deployability of the solutions. Others required installing some software at the user level without the need for recompiling the kernel \cite{DBAS,DNIS}. This installation adds a new layer, which takes the responsibility of utilizing the available interfaces. Rodriguez et al. \cite{MAR} and Hasegawa et al. \cite{hasegawa2007} avoid updating client devices by implementing their solution in the router. It is only required to configure the client device by setting the default DNS server to be the router itself.

\textit{\textbf{- Deployment Attempts:}}
Throughout the last decades, there were some attempts to deploy bandwidth aggregation systems in the Internet. For instance, Linux implemented true or trivial link equalizer (TEQL) \cite{li2006performance}, a link bonding technique to enable users to utilize their multiple interfaces. Unfortunately, TEQL is only suitable for directly connecting devices with multiple homogeneous interfaces as well as connecting a device to a gateway with multiple homogeneous links. Therefore, the nature of having multiple heterogeneous interfaces at mobile devices led to performance degradation while using TEQL. Hence, TEQL is not enabled by default in the current versions of Linux.

Recently, with the availability of multi-homed mobile devices, their nature of having heterogeneous interfaces, and the increasing user demand for bandwidth,  IETF released the multi-path TCP standard \cite{AppleTCP} which is adopted by Apple on their iOS 7. Although apple successfully deployed this multi-path TCP protocol in millions of mobile devices in the first few weeks, it is not considered a successful deployment yet since it works only with Apple servers while using Siri application. Although Apple demonstrated the ability of updating millions of client devices in few weeks, the multi-path TCP protocol deployment attempt relies on (1) the willingness of Internet server operators to deploy this standardized version of multi-path TCP at their side, and (2) the willingness of Internet middle-boxes operators to upgrade their middle-boxes to deal with this new protocol \cite{raiciu2012hard}.

\begin{figure}[!t]
  \centering
  \includegraphics[width=\columnwidth]{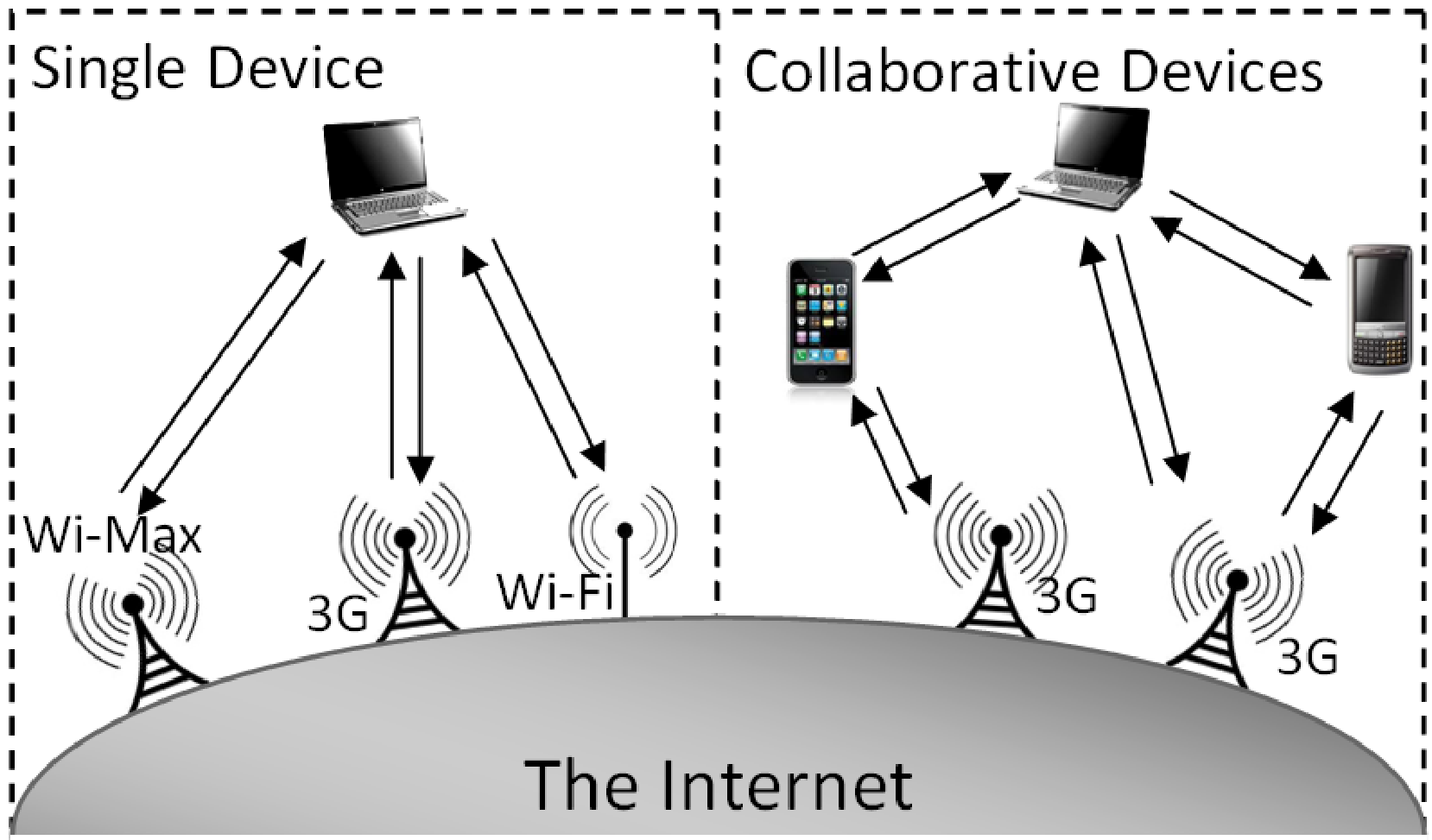}\\
  \caption{Single versus collaborative device scheduling.}\label{F1-2}
\end{figure}

\subsubsection{Utilizing Middle-boxes}
Although relying on middle-boxes such as proxy servers increases the deployment cost, the existence of content distribution networks demonstrates the acceptance of this cost if it results in tremendous performance enhancement. To date, researchers focus on using middle-boxes to enable bandwidth aggregation and address many challenges such as scheduling and interface characteristics estimation. However, they overlook many challenges introduced by using middle-boxes and potential usage of these middle-boxes such as 1) scalability and fault tolerance , 2) proxy placement, 3) traffic redirection, and 4) performance optimization.

\textit{\textbf{- Scalability and Fault Tolerance:}}
With the widespread deployment and adoption of middle-box based solutions, the available bandwidth and the computation power at these middle-boxes would become the performance bottleneck leading to performance degradation at the client side. Hence, to handle many devices, researchers should give sufficient attention to the scalability of their approaches and their design. In addition, efficiently handling middle-box failures is critical to avoid decreasing the quality of user experience.

\textit{\textbf{- Proxy Placement:}} Since the client Internet traffic will pass through the middle-boxes, the location of these middle-boxes can significantly impact performance. Optimally, these middle-boxes should be placed on the route between clients and servers and should be connected to the servers via high speed links. Such placement avoids significant increase in client-server latency but it is almost infeasible because of the widespread deployment of Internet servers with whom a single client communicates at any point in time. Therefore, we argue that placing middle-boxes at the edge of Internet close to clients minimizes the latency overhead. In contrast, placing these middle-boxes at the Internet edge increase the deployment cost due to the need of more middle-boxes to cover all the Internet edge networks. Hence, researchers should handle the performance-cost tradeoff introduced by their middle-box placement techniques.

\textit{\textbf{- Traffic Redirection:}} Seamless traffic redirection mechanisms is a critical component for distributing the load across the different middle-boxes and handling faults. Designing an efficient traffic redirection mechanism is considered a tremendous challenge because it should consider multiple aspects such as 1) maintaining end-clients and end-servers communication status, 2) maintaining the client interface information, and  3) avoid performance degradation while redirecting the traffic.

\textit{\textbf{- Performance Optimization:}} Using middle-boxes enables many performance optimization techniques that significantly enhance the performance. For instance, caching the server contents can significantly enhance the response time. On the other hand, using middle-boxes enables adopting many communication protocols that are suitable for the client and can avoid performance degradation due to interface characteristics (e.g. loss in wireless environments). This enables data bundling and compression at the client or the middle-box to make best use of the client's limited bandwidth. Furthermore, It enables utilizing opportunistic networks like Wifi in bus stations.

\subsubsection{Enabling Client Collaboration}
Although most bandwidth aggregation work focuses on utilizing the interfaces on a single device, there are a few attempts that target exploiting the bandwidth available on interfaces in neighboring devices \cite{TMCMCC,QAWBA,combine,COBA,OSCAR,OSCAR-MASS,GMCC-Coop} (Figure~\ref{F1-2}).  The motivation behind this collaborative approach is the increased mobile device density, the availability of high speed ad-hoc links amongst them and the lack of continuous, reliable, high bandwidth, and cheap Internet connectivity \cite{CollMotivation}. In such cases, each device has at least two different network interfaces. The first interface is directly connected to an expensive limited bandwidth to Internet while the other interface can be used in collaboration between the different devices. This collaboration introduces a new set of challenges including 1) neighbor discovery, 2) user incentives, 3) security issues, 4) data scheduling, and 5) sharing caches.

\textit{\textbf{- Neighbor Discovery:}}
One of the most important features of a collaborative bandwidth aggregation system is the ability of a device to discover its neighbors and gather essential information about them (e.g their ability to share bandwidth and how much they are willing to share). Available solutions investigate the neighbor discovery problem in different ways. QAWBA\cite{QAWBA} implements a k-hop neighbor discovery protocol in which a QAWBA node discovers an Internet sharing node within k-hops. Other approaches discover neighboring devices using a proxy server that handles device collaboration \cite{TMCMCC,PRISM}.

Unfortunately, developed approaches to date discover neighbors and return their shared bandwidth assuming that this bandwidth comes from a single interface. This assumption is not always true since the neighboring device may be equipped with multiple interfaces connected to the Internet and already sharing these different connections that have different characteristics. Recently, OSCAR \cite{OSCAR,OSCAR-MASS} took initial step towards solving this problem by enabling devices to return their different connectivity options that they are willing to share. Although OSCAR enables devices to share bandwidth that is shared with them from other devices, in many cases, OSCAR nodes fail in discovering shared bandwidths that are reachable through multiple hops. On the other hand, using a proxy to discover neighboring devices and handle collaboration is not efficient since it introduces communication overhead and synchronization problems between clients and the proxy. It also limits the widespread deployment of the system since proxies can be a bottleneck in this case.

\textit{\textbf{- User Incentives:}}
It is critical for a collaborative bandwidth aggregation system to be equipped with an incentive system to encourage users to share their bandwidth. Creating and integrating effective incentive systems with bandwidth aggregation solutions can only lead to the popularity and the widespread deployment of the system. Borrowing techniques and ideas from other incentive system solutions is a good way to start. These systems can be categorized into three categories: (1) game-theoretic based systems; (2) reputation-based systems; and (3) credit-based systems.

Game-theoretic based incentive systems rely on the rationality of the game players. These approaches designs a game in which the collaborating nodes will not gain or even lose if they try to cheat or do not collaborate \cite{game1,game2}. They assume that all the nodes have global knowledge about the game status and can interact accordingly. Such assumption creates an obstacle for adopting this type of incentive schemes in the context of large scale collaborative bandwidth aggregation systems as maintaining and spreading this information will introduce a huge overhead.

In reputation-based systems, each node builds a reputation by serving other nodes in order to be served in the future. In such systems, each node carries the overhead of monitoring its neighbors since they most probably will collaborate with them later. It is also responsible for spreading gathered observation regarding neighbors through the network to enable the other nodes determine their reputation levels and act accordingly. Sharma et al. \cite{TMCMCC} used this approach to provide incentives for their small collaborative community. This approach does not scale, however, and prevents such systems from being fully deployed over the Internet.

On the other hand, credit-based systems are suitable for large scale networks since they usually rely on a trusted third party that maintains credits for the communicating nodes \cite{credit1,credit2}. These nodes usually collaborate with each other and every node pays for the service it requests. The collaborating nodes who offer the service usually gain credit which they use for their own benefit. OSCAR \cite{OSCAR,OSCAR-MASS} uses this approach to provide incentive for its users. This advantage makes this kind of incentive system best suitable for widely used collaborative bandwidth aggregation systems.

\textit{\textbf{- Security:}}
Security is one of the most important challenges in a collaborating environment. Selfish node behavior may drive them towards cheating in order to exploit the collaborating nodes connectivity without sharing their own resources. More importantly, nodes may eavesdrop, alter, or maliciously compromise relayed data. Connections, if not properly authenticated can be hijacked and users may be impersonated. These kinds of challenges should be efficiently addressed and security-based solutions need to be adopted and tailored for bandwidth aggregation environments.

\textit{\textbf{- Scheduling:}}
Scheduling on neighboring devices interfaces adds a lot of challenges including monitoring their gain as well as implementing deployable and seamless relaying. Schedulers need to address other metrics like incentive cost, energy consumption across devices, and fair throughput for all devices. Schedulers will also need to decide on which interfaces to use for scanning and sharing connectivity.

\textit{\textbf{- Sharing Caches:}}
To enhance performance and achieve better utilization of the available bandwidth of Internet connections, collaborating devices can share their caches together. For instance, sharing DNS caches is an approach to increase the responsiveness of Internet applications. On the other hand, sharing HTML-5 caches may significantly enhance the performance of HTML-5 applications and online games. It avoids wasting the bandwidth of Internet connections in transferring redundant data.

\section{Tangential Research Areas}
\label{RW}
\begin{figure}[!t]
  \centering
  \includegraphics[width=\columnwidth]{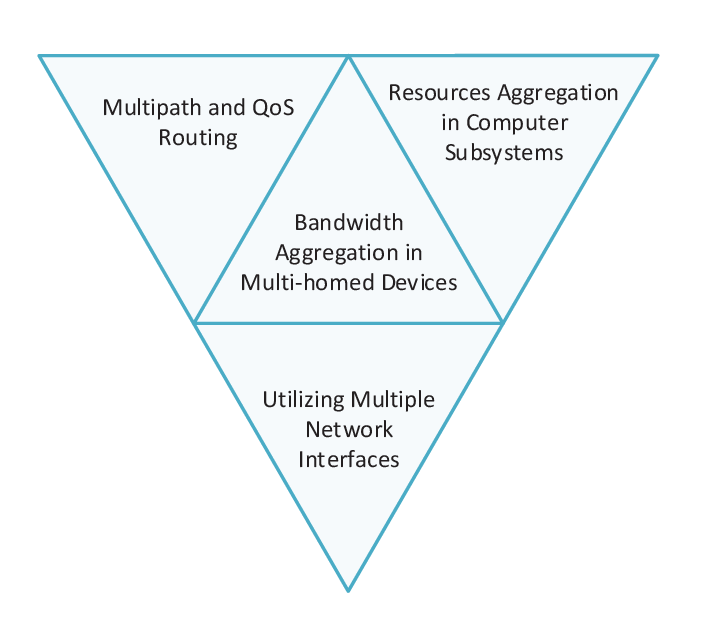}\\
  \caption{Tangential research areas to the bandwidth aggregation on multi-homed devices problem.}\label{RelatedWork}
\end{figure}

There are three tangential research areas close to the multi-interface
bandwidth aggregation problem as shown in (Figure \ref{RelatedWork}): (1)
multi-path routing, (2) resources aggregation in computer sub-systems, and (3)
utilizing the availability of multiple network interfaces for non-bandwidth
aggregation purposes.

\subsection{Multi-path Routing}
Multi-path routing refers to using multiple routes between the source and the
destination \cite{survey,qadir2015exploiting}. The multiple routes are selected
to go through multiple neighbors, which is usually performed for reliability
and security reasons. Data can be replicated on the different paths
\cite{redundant}, distributed on multiple paths
\cite{FZR,BGR,tsirigos2001,cetinkaya2007improving}, encoded to enhance the
reliability \cite{coding}, or sent using a single path, while  maintaining
other paths as backup \cite{backup}.

Multi-path QoS routing inherits all the characteristics from the multi-path
routing while making use of the available multiple paths to grant a certain QoS
level to the different streams \cite{wang2007interference,an2005cluster}. These
approaches use multiple metrics to quantify the QoS given to the streams such
as delay and bandwidth.

\subsection{Resource Aggregation in Computer Sub-systems}
Aggregating available resources to obtain higher performance is a known problem
in traditional computer systems \cite{LS2,LS3,FemtoCloud,COSMOS}. Data stripping 
is the key aspect in the Redundant Arrays of Inexpensive Disks (RAID) architecture
\cite{LS2} for disk sub-systems.  Traw et al. provide an overview on resource
aggregation in network sub-systems \cite{LS3}. They introduce an evaluation
criteria to measure the benefits of the aggregated resources in terms of
latency, buffering requirements, skew tolerance, scalability, complexity, and
finally the maximum aggregate bandwidth which can be supported. These network
sub-systems handle communication between different processors, processor and
memory, as well as different hosts. They take the TCP/IP protocol stack network
sub-system as a case study. However, they did not address the implementation
layer, scheduling, nor evaluation.

\subsection{Utilizing Multiple Network Interfaces}
Utilizing the available multiple network interfaces has been an active research
area during the last several years. In this survey we address utilizing them
for bandwidth aggregation. However, there exists a large body of research work
that utilizes these interfaces to achieve other goals.

Some researchers utilize the available interfaces in order to minimize using
the highly loaded cellular networks through mobile data offloading
\cite{de2011ip,aijaz2013survey}. For instance, the Wiffler system
opportunistically offloads data over WiFi to minimize the use of these cellular
networks when they become heavily loaded \cite{Intentional3}. Wiffler was
developed since the currently used techniques aimed to encourage users minimize
cellular network load, such as imposing a limit of 5 GB per month or educating
users on responsible network access, are deemed ineffective and insufficient. 

Other researchers exploit multiple interfaces in order to minimize energy
consumption. Some of them use the interface with low energy consumption to wake
up other interfaces \cite{Wiff20,Wiff4,Intentional2}. Johansson et al.
\cite{Intentional17} leverage these interfaces to reduce the energy consumption
as well. They show that Bluetooth radios are often preferable to IEEE 802.11xB
for short-range communication.

Others utilize these interfaces to handle mobility and overcome the wireless
challenges \cite{Intentional2}. They utilize identical wireless interfaces in
order to increase the seamlessness of the wireless access point handoff. They
propose techniques that tolerate the wireless link problems and avoid drawbacks
of random backoff. They propose using one interface to control the media and
prepare the schedule which other interfaces follow to transmit its data. They
also propose increasing the communication capacity by using multiple wireless
interfaces tuned to different channels.

Finally, some researchers utilize the available interfaces in the context of
cognitive radio and mesh networks. In cognitive radio networks, they leverage
these interfaces to avoid interfering with primacy users and to create better
spectrum opportunities \cite{Cognitive1,Cognitive2,DZP}. In addition, some
interfaces can be dedicated and used for control traffic \cite{LAUNCH}. In the
context of mesh networks, Draves et al. show how the overall throughput can be
increased for multi-radio nodes by dynamically choosing the ``best'' outbound
link when forwarding a given packet \cite{Intentional10}.

\section{Conclusion}
\label{con}
In this paper we have surveyed the most prominent solutions proposed for addressing bandwidth aggregation problems in multi-homed devices. We have discussed the problem, examined and analyzed the proposed research, and showed its tangential areas. We analyzed the different features of the problem solution and discussed how each solution implemented each of these features. Finally, we have analyzed the various evolution trends and discussed potential open challenges we believe researchers can pay attention to.

\bibliographystyle{elsarticle-num}
\bibliography{Survey_v6}
\newpage
\end{document}